\documentclass[%
 aip,
rsi,%
 amsmath,amssymb,
 reprint,%
longbibliography,
]{revtex4-1}

\usepackage{xcolor, soul}
\sethlcolor{yellow}
\usepackage{textgreek}
\usepackage{graphicx}
\usepackage{dcolumn}
\usepackage{bm}

\begin{document}

\preprint{AIP/123-QED}

\title[Chiral metasurfaces formed by 3D-printed square helices]{Chiral metasurfaces formed by 3D-printed square helices: A flexible tool to manipulate wave polarization}
\author{Shengzhe~Wu}
\affiliation{College of Physics, Jilin University, 2699 Qianjin Street, Changchun 130012, China}  
\author{Vladimir~V.~Yachin}
\affiliation{Institute of Radio Astronomy of National Academy of Sciences of Ukraine, 4 Mystetstv Street, Kharkiv 61002, Ukraine}%
\affiliation{V. N. Karazin Kharkiv National University, 4 Svobody Square, Kharkiv, 61022, Ukraine}
\author{Vitalii I. Shcherbinin} 
\affiliation{National Science Center `Kharkiv Institute of Physics and Technology', National Academy of Sciences of Ukraine, 1, Akademicheskaya Street, Kharkiv 61108, Ukraine}
\affiliation{International Center of Future Science, State Key Laboratory of Integrated Optoelectronics, College of Electronic Science and Engineering, Jilin University, 2699 Qianjin Street, Changchun 130012, China}
\author{Vladimir~R.~Tuz}
 \email{tvr@jlu.edu.cn; tvr@rian.kharkov.ua}
\affiliation{Institute of Radio Astronomy of National Academy of Sciences of Ukraine, 4 Mystetstv Street, Kharkiv 61002, Ukraine} 
\affiliation{International Center of Future Science, State Key Laboratory of Integrated Optoelectronics, College of Electronic Science and Engineering, Jilin University, 2699 Qianjin Street, Changchun 130012, China}

\date{\today}

\begin{abstract}
The transmission of linearly and circularly polarized waves are studied both theoretically and experimentally for chiral metasurfaces formed by arrays of metallic square helices. The helical particles of the metasurfaces are constructed of rectangular bars manufactured by direct three-dimensional printing in solid metal. The transmittance of the metasurface is found to depend critically on the number of bars forming the square helical particles. In the case of an even number of bars, the chiral metasurface exhibits identical co-polarized transmittance of orthogonal linearly polarized waves, which are characterized by a dual-band asymmetric transmission. For an odd number of bars, the metasurface provides the same cross-polarization conversion for any polarization orientation of the incident field and thus serves as a polarization-independent twist polarizer. Finally, the transmittance of this polarizer is investigated with respect to the dimensions of the square helices. The investigated chiral metasurfaces are characterized by strong broadband circular dichroism regardless of the number of bars in the helical particles. The wide variety of transmission properties observed in the metasurfaces makes them particularly attractive for use in polarization conversion and separation devices.
\end{abstract}


\maketitle

\section{\label{sec:intro}Introduction}

Chiral media are a class of optically active substances.\cite{Barron_book_2004} Owing to their helical structures, these media are characterized by intrinsic left- or right-handedness. Chiral media provide the polarization transformation of propagating waves, which can experience the effects of circular dichroism, asymmetric transmission (polarization conversion dichroism), and optical rotatory dispersion.\cite{Hecht_2016} Recent advances in polymer science and nanofabrication techniques have generated renewed interest in chiral media, which can currently be realized as artificial structures (metamaterials).\cite{Svirko_ApplPhysLett_2003, Prosvirnin_PhysRevE_2005, Fedotov_PhysRevLett_2006, Prosvirnin_JOpt_2009, Plum_PhysRevB_2009, plum2010asymmetric, Engelbrecht_ApplPhysLett_2010, Shi_ApplPhysLett_2013, Shadrivov_ApplPhyLett_2013, Polevoy_2013, Song_JApplPhys_2016, Zhang_PhysRevLett_2009, Fang_JApplPhys_2017} Compared to natural optically active media, chiral metamaterials can produce much stronger chiral-dependent effects and are therefore being explored for expanded applications in science and technology (see the progress reports in Refs.~\onlinecite{Ozbay_JOpt_2013, Valev_AdvMater_2013, Hentschel_AAAS_2017, Collins_AdvOptMat_2017}).

Among particles used to construct chiral metamaterials, spirals (or helices) are intrinsically chiral (true) three-dimensional (3D) objects (a 3D chiral object retains the same handedness regardless of its rotation\cite{Arteaga_OptExpress_2016}). Using modern lithography techniques and direct laser writing, 3D chiral structures can be realized on the micron scale.\cite{Gansel_Science_2009, Wegener_AdvMater_2009,  Gansel_ApplPhysLett_2012, Sakellari_AdvOptMater_2017, Tsutsumi2017, Golod_AdvOptMater_2018, Zhu_LightSciAppl_2018, Gorkunov_SciRep_2018} Despite great progress in these techniques, the fabrication of 3D structures is still challenging and requires expensive equipment and materials.

A straightforward way to resolve this problem is to replace the helix with simple individual elements (e.g., bars \cite{Wu_PhysRevApplied_2014, Park_NaCommun_2014, Fasold_acsphotonics_2018}) arranged in a twisted structure. The multilayered arrangement of several elements can be considered as the first winding of a helix. This design strategy has three benefits:\cite{Hentschel_ACSNano_2012} (i) the fabrication of such structures is possible by either self-assembly or multilayer electron-beam lithography; (ii) either capacitive coupling (near-field coupling) or conductive coupling (for touching elements) between the elements forming the helix is possible, allowing the flexible tuning of the corresponding modes;\cite{Slaughter_ACSNano_2010} and (iii) the spectral regime in which the chiral optical response occurs can be nearly arbitrarily shifted by changing either or both the size and material of the elements. In such multilayered structures, spectral tunability and broadband operation become possible along with strong circular dichroism.\cite{Hentschel_ACSNano_2012} Furthermore, several widely used approaches are now available to characterize the multilayered arrangement of elements in a chiral medium.\cite{Svirko_ApplPhysLett_2001, Menzel_PhysRevA_2010, Pavlov_2013, Qu_AnnPhys_2018, Zhang_JOptSocAmB_2018, Sperrhake_OptExpress_2019}

Another promising technique for producing true 3D chiral structures is 3D printing.\cite{Headland_OptExpress_2016, Camposeo_AdvOptMater_2019} In 3D printing, a structure is built up in a layered fashion. Remarkably, in modern 3D printers, the selective laser melting makes it possible to perform printing directly in solid metals. In this process, the laser melts a certain portion of the powder metal and then fuses it and cools it down to form the solid portion of the relevant layer. This technique allows the realization of complex and high-precision metallic structures with resolutions on the order of a few tens to a few hundreds of micrometers. The beneficial properties of the resulting structures make them particularly attractive for the manipulation of microwave and terahertz waves with wavelengths between 30 \textmu m and 1 mm. 

In this paper, we extend the results of our letter,\cite{Wu_OL_2019} which deals with a true 3D chiral metasurface whose particles (square helices) are constructed from six rectangular bars stacked together in a multilayered fashion. Such a metasurface was found to exhibit a dual-band asymmetric transmission and strong broadband circular dichroism. However, it is not clear how general the findings of Ref.~\onlinecite{Wu_OL_2019} are. For this reason, in this paper, we present a systematic theoretical and experimental investigation of true 3D chiral metasurfaces formed by square helices with different designs. First, the rigorous theoretical method introduced in Ref.~\onlinecite{Yachin_JOSAA_2007} is generalized to multilayered structures with the aim of calculating the transmission characteristics of a metasurface constructed from a set of double-periodic slabs. Next, using the developed method, we study the polarization transformations of the linearly and circularly polarized waves for chiral metasurfaces made of helices with different (even and odd) numbers of bars. For these metasurfaces, chiral-dependent effects (strong asymmetrical transmission, highly efficient cross-polarization conversion, and broadband circular dichroism) are demonstrated. In addition, we investigate the structural parameters to ensure the optimal optical response of the metasurface. Finally, all distinctive features of the proposed metasurfaces are verified experimentally  using 3D-printed prototypes manufactured for quasi-optical (microwave) measurements.

\section{\label{sec:theory}	Theory}

\begin{figure}[t!]
\centering
\includegraphics[width=1\linewidth]{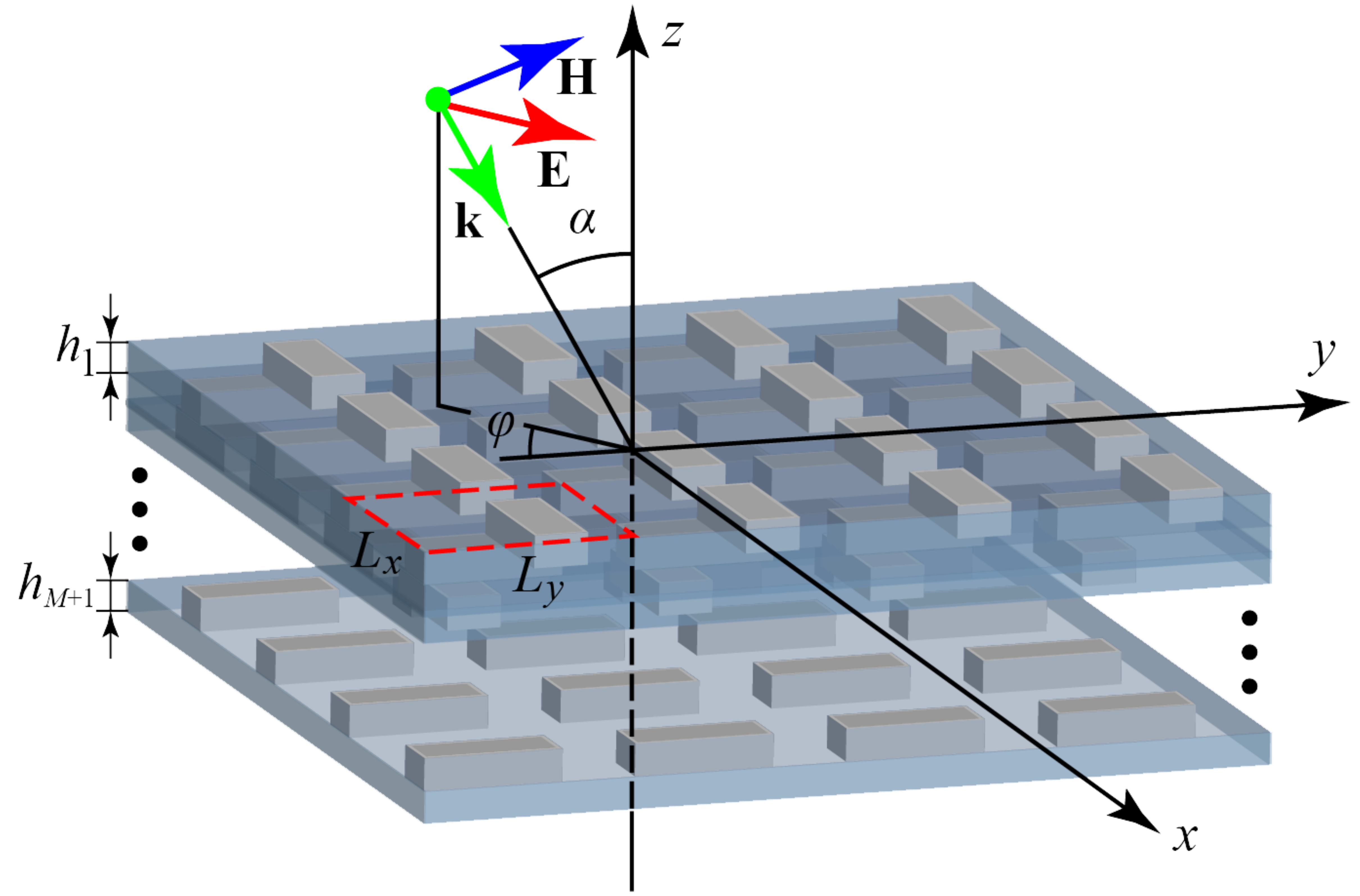}
\caption{Schematic of a metasurface composed of a multilayered system of bars.}
\label{fig:scheme}
\end{figure} 

The scattering problem of interest is formulated in a general form with the geometry shown in Fig.~\ref{fig:scheme}. A linearly polarized plane wave of unit amplitude is incident on a ($M+1$)-layered medium at the angle $\alpha$ with respect to the $z$-axis. The incident plane wave is a superposition of transverse electric (TE) and transverse magnetic (TM) waves. For the TE (TM) wave, the electric (magnetic) field vector $\mathbf{E}_0(\mathbf{r})$ [$\mathbf{H}_0(\mathbf{r})$] belongs to the $x$--$y$ plane and is rotated by the angle $\varphi$ with respect to the $x$-axis. Each layer of the ($M+1$)-layered medium is a double-periodic slab whose translation unit cell is composed of dielectric particles buried in a substrate. The substrate and particle materials are characterized by relative permittivities $\varepsilon_1$ and $\varepsilon_2$  and permeabilities $\mu_1$ and $\mu_2$, respectively. The layers have thicknesses $h_k$ ($k = 1, 2, \ldots, M+1$). The dimensions of the unit cell are $L_x$ and $L_y$ along the $x$- and $y$-axes, respectively. The structure is surrounded by an ambient space with material parameters $\varepsilon_0$ and $\mu_0$.

First, we consider a single-layered structure with thickness $h_1$. The layer is double-periodic in the transverse direction ($x$--$y$ plane), and its relative permittivity and permeability can be written in the following form:
\begin{equation} 
\varepsilon(\mathbf{r_{\bot}}) = \varepsilon(x, y), \quad \mu(\mathbf{r_{\bot }}) = \mu (x, y), \label{EQ__1_}
\end{equation} 
with periodic conditions
\begin{equation}
\begin{split}
\varepsilon (x + L_x, y + L_y) &= \varepsilon (x, y), \\ 
\mu(x + L_x, y + L_y) &= \mu(x, y). 
\end{split}
\label{EQ__2_} 
\end{equation} 

Using the approach of the equivalent electric and magnetic polarization currents, we can write the scattered electric and magnetic fields as\cite{Yachin_JOSAA_2007}
\begin{equation} 
\begin{split}
\label{EQ__6_}
\mathbf{E(\mathbf{r})}=\mathbf{E}_0(\mathbf{r}) &+ \frac{1}{4\pi}\left(\nabla \nabla + k^2\right)\int_V G (\mathbf{r}, \mathbf{r'})\mathbf{J}^e(\mathbf{r'})d\mathbf{r'}
\\
&+ \frac{ik}{4\pi} {\rm \nabla }\times \int_V G (\mathbf{r}, \mathbf{r'})\mathbf{J}^h(\mathbf{r'})d\mathbf{r'},
\end{split}
\end{equation} 
\begin{equation}
\begin{split}
\label{EQ__7_} 
\mathbf{H}(\mathbf{r})=\mathbf{H}_0(\mathbf{r}) &+ \frac{1}{4\pi}\left(\nabla \nabla + k^2\right)\int_V G(\mathbf{r}, \mathbf{r'})\mathbf{J}^h(\mathbf{r'})d\mathbf{r'}
\\
&- \frac{ik}{4\pi} {\rm \nabla }\times \int_V G(\mathbf{r},\mathbf{r'})\mathbf{J}^e(\mathbf{r'})d\mathbf{r'},
\end{split}
\end{equation} 
where $G(\mathbf{r},\mathbf{r'})=\exp (ik\left|\mathbf{r}-\mathbf{r'}\right|)/\left|\mathbf{r}-\mathbf{r'}\right|$ is the Green's function in free space, $k=\omega \sqrt{\varepsilon_0 \mu_0}$ is the wavenumber in free space, $V$ is the volume of the double-periodic layer, and $\mathbf{r}$ and $\mathbf{r'}$ denote the positions of the observation and source points, respectively. The equivalent electric [$\mathbf{J}^e(\mathbf{r})$] and magnetic [$\mathbf{J}^h(\mathbf{r})$] polarization currents  are found from the following equations:
\begin{equation}
\begin{split}
\label{EQ__4_} 
\mathbf{J}^e(\mathbf{r}) =& \left[\varepsilon (\mathbf{r_{\bot}}) - 1\right]\mathbf{E}(\mathbf{r}), \\
\mathbf{J}^h(\mathbf{r}) =& \left[\mu (\mathbf{r_{\bot}}) - 1\right]\mathbf{H}(\mathbf{r}).  
\end{split}
\end{equation} 

To solve Eqs.~(\ref{EQ__6_})--(\ref{EQ__4_}), we suppose that the observation point $\mathbf{r}$ is located inside the layer and introduce the integral functionals $\mathbf{I}^e(\mathbf{r})$ and $\mathbf{I}^h(\mathbf{r})$ of the form:
\begin{equation}
\label{EQ__14_}
\mathbf{I}^{e(h)}(\mathbf{r}) = -\frac{i}{2\pi} \int_V G(\mathbf{r}, \mathbf{r'})\mathbf{J}^{e(h)}(\mathbf{r'})d\mathbf{r'}.     
\end{equation}     

With Eq.~(\ref{EQ__14_}), the integro-differential Eqs. (\ref{EQ__6_}) and (\ref{EQ__7_}) can be reduced to the differential equations for the functionals $\mathbf{I}^e(\mathbf{r})$ and $\mathbf{I}^h(\mathbf{r})$.

Taking into account the double periodicity of the layer, the integral functionals $\mathbf{I}^{e(h)} (\mathbf{r})$ can be written as follows:
\begin{equation} 
\label{EQ__18_}
\mathbf{I}^{e(h)}(\mathbf{r}) = \sum_{p=-\infty}^{\infty} \sum_{q=-\infty}^{\infty} I_{pq}^{e(h)}(z)e^{ik_{x_1, p} x_1} e^{ik_{y_1, q} y_1},    
\end{equation} 
where $k_{x, p} = k_x +2p\pi/L_x$, $k_{y, q} = k_y + 2q\pi/L_y$, $k_x = -k\sin\alpha \sin\varphi$, and $k_y = k\sin\alpha \sin\varphi$. 

Using Galerkin's method, from Eqs.~(\ref{EQ__6_}) and (\ref{EQ__7_}), we derive a matrix differential equation with constant coefficients. In the case of an incident TE wave, the general solution of the matrix equation takes the following form:
\begin{equation}
\begin{split}
\label{EQ__38_} 
I_{x(y),pq}^e(z) = \sum_{m=1}^{2P}\left(c_m^+ e^{i\chi_m z}\right. &+ \left.c_m^- e^{-i\chi_m z}\right) W_{x(y), pq, m}^e\\ 
&+ \frac{2i}{k^2} E_{0x(0y)} e^{ik_z z},  
\end{split} 
\end{equation} 
\begin{equation} 
\label{EQ__39_} 
I_{z,pq}^e(z) = \sum_{m=1}^{2P}\left(c_m^+ e^{i\chi_m z} + c_m^- e^{-i\chi_m z}\right) W_{z, pq, m}^e,  
\end{equation} 
\begin{equation}
\label{EQ__40_} 
I_{\nu ,pq}^h(z) = \sum_{m=1}^{2P}\left(c_m^+ e^{i\chi_m z} +c_m^- e^{-i\chi_m z}\right) W_{\nu, pq, m}^h,
\end{equation} 
where $P=(2N_x + 1)\times (2N_y + 1)$, $\{c_m^\pm\}$ are the unknown constants, $N_x$ and $N_y$ are the numbers of the considered Floquet modes propagated along the $x$- and $y$-axes, respectively,  $\chi_m$ and $W_{\nu, pq, m}^{e(h)}$ are the $m$-th eigenvalue and eigenvector of the characteristic equation derived from the homogeneous part of the obtained matrix equation, respectively, and $\nu =x,y,z$. The solution for the incident TM wave with magnetic field $\mathbf{H}_0(\mathbf{r})=\{H_{0x}(\mathbf{r}),H_{0y}(\mathbf{r}),0\}$ can be readily deduced from that for the TE wave using the duality of electric and magnetic fields.

Next, we consider ($M+1$)-layered double-periodic structures. In this case, we can apply the above-described procedure to derive functionals for each layer independently. Then, using the boundary conditions on the interfaces between adjacent layers, we obtain the following system of equations for unknown constants:
\begin{widetext}
\begin{equation}
\label{eq_1}
\begin{pmatrix} 
{-F_1 c_1^+} & {-Z_1 c_1^-} & {0} & {0} & {0} & {0}  \\
{Z_1 [e^{\lambda_1 h_1} ]c_1^+} & {F_1 [e^{-\lambda_1 h_1}]c_1^-} & {-Z_2 c_2^+} & {-F_2 c_2^-} & {0} & {0}  \\ 
{F_1 [e^{\lambda_1 h_1}]c_1^+} & {Z_1 [e^{-\lambda_1 h_1}]c_1^-} & {-F_2 c_2^+} & {-Z_2 c_2^-} & {0} & {0} \\ 
{0} & {0} & {Z_2 [e^{\lambda_2 h_2}]c_2^+} & {F_2 [e^{-\lambda_2 h_2}]c_2^-} & {-Z_3 c_3^+} & {-F_3 c_3^-} \\
{0} & {0} & {F_2 [e^{\lambda_2 h_2}]c_2^+} & {Z_2 [e^{-\lambda_2 h_2}]c_2^-} & {-F_3 c_3^+} & {-Z_3 c_3^-} \\ 
{\vdots} & {\vdots} & {\vdots} & {\vdots} & {\vdots} & {\vdots} \\
{Z_M [e^{\lambda_M h_M}]c_M^+} & {F_M [e^{-\lambda_M h_M}]c_M^-} & {-Z_{M+1} c_{M+1}^+} & {-F_{M+1} c_{M+1}^-} & {0} & {0} \\ 
{F_M [e^{\lambda_M h_M} ]c_M^+} & {Z_M [e^{-\lambda_M h_M}]c_M^-} & {-F_{M+1} c_{M+1}^+} & {-Z_{M+1} c_{M+1}^-} & {0} & {0} \\
{0} & {0} & {Z_{M+1} [e^{\lambda_{M+1} h_{M+1}}]c_{M+1}^+} & {F_{M+1} [e^{-\lambda_{M+1} h_{M+1}}]c_{M+1}^-} & {0} & {0} 
\end{pmatrix} =
\begin{pmatrix} 
{b} \\
{0} \\
{0} \\
{0} \\
{0} \\
{\vdots} \\
{0} \\
{0} \\
{0}
\end{pmatrix},
\end{equation}
\end{widetext}
where $\exp(\pm\lambda_m h_m)$ are the diagonal matrices related to the eigenvalues of the $m$-th layer of the structure, and $h_m$ is the thickness of the $m$-th layer (the explicit form of the matrices $Z_m$ and $F_m$, which are derived from the integral form of the Maxwell equations, can be found in Ref.~\onlinecite{Yachin_JOSAA_2007}).

Using a finite number of straightforward mathematical operations, the unknowns $c_{m}$ can be found from the system of equations (\ref{eq_1}). Among these unknowns, the constants $c_1^\pm$ and $c_{M+1}^\pm$ completely determine the reflection and transmission coefficients of the ($M + 1$)-layered structure, respectively. An efficient way to find $c_1$ and $c_{M+1}$ is to use the recursive scheme.\cite{Li_JOSAA_1996, Moharam_JOSAA_95, Li_JOSAA_2003} To find $c_{M+1}^\pm$, we first express the unknown $c_1^-$ of the first block row of the matrix (\ref{eq_1}) in terms of $\bar{c}_1^+$, where $\bar{c}_1^+=[e^{\lambda_1 h_1} ]c_1^+$. Next, we express $\bar{c}_1^+$ in terms of $c_2^-$ and exclude it from the second and third block lines. Using such a recursive scheme for $c_m^-$ $(m = 2, \ldots, M+1)$, we finally arrive at the following expressions for $c_{M+1}^-$:
\begin{eqnarray}
\label{eq_2}
c_{M+1}^- =\left(B_{M+1} - A_{M+1} \left[e^{-\lambda_{M+1} h_{M+1}}\right]Z_{M+1}^{-1} F_{M+1}\right) \nonumber\\
\times  \left[e^{-\lambda_{M+1} h_{M+1}} \right]\left(\prod_{m=2}^M \bar{B}_m \left[e^{-\lambda_m h_m}\right]B_m^{-1} \right)\nonumber\\
\times \bar{B}_1 \left[e^{-\lambda_1 h_1} \right]Z_1^{-1} b,\qquad 
\end{eqnarray}
where $$X_1 = Z_1 - F_1 \left[e^{-\lambda_1 h_1}\right]Z_1^{-1} F_1 \left[e^{-\lambda_1 h_1}\right],$$ 
$$Y_1 = F_1 - Z_1 \left[e^{-\lambda_1 h_1}\right]F_1^{-1} Z_1 \left[e^{-\lambda _1 h_1}\right],$$ 
$$A_m = Y_{m-1} X_{m-1}^{-1} Z_m - F_m,$$ 
$$B_m = Y_{m-1} X_{m-1}^{-1} F_m - Z_m,$$ 
$$X_m = Z_m - F_m \left[e^{-\lambda_m h_m}\right]B_m^{-1} A_m \left[e^{-\lambda_m h_m}\right],$$ 
$$Y_m =F_m - Z_m\left[e^{-\lambda_m h_m}\right]B_m^{-1} A_m\left[e^{-\lambda_m h_m}\right],$$ 
$$\bar{A}_m = Y_m X_m^{-1} Z_m - F_m,$$ 
$$\bar{B}_m = Y_m X_m^{-1} F_m - Z_m,$$ 
and $\bar{c}_{M+1}^+ =-Z_{M+1}^{-1} F_{M+1}\left[e^{-\lambda_{M+1} h_{M+1}}\right]c_{M+1}^-$.

The constants $c_{M+1}^{\pm }$ determine the transmission coefficient of the plane wave scattered from the double-periodic multilayered structure, where $c_{M+1}^+=[e^{-\lambda_{M+1} h_{M+1}} ]\bar{c}_{M+1}^+$. The expressions for $c_1^\pm$, which describe the wave reflection from the structure, can be found using a similar recursive scheme. Compared to the recursive $S$-matrix method,\cite{Li_JOSAA_2003} our approach requires half as many matrix inversions to calculate either $c_1^\pm$ or $c_{M+1}^\pm$ and is therefore faster and more accurate. 

We now use the above-described theory to simulate wave transmission through a multilayered structure composed of metallic particles. To do this, we assume that the layer permittivity and permeability satisfy the conditions $|\varepsilon| \gg 1$ and $\varepsilon\mu\simeq 1$. Under these conditions, the constituents (rectangular bars) of the layers behave like those made of perfect electric conductors. 

It should be noted that our approach has an advantage over the widely known method of moments, which requires the 3D discretization of the scattering problem for volumetric structures. Specifically, compared to commercial electromagnetic solvers (e.g., HFSS and Comsol Multiphysics), our algorithm for solving the scattering problem is superior in the case of multilayered structures. In our approach, discretization is carried out in layers (along the structure) followed by the two-dimensional discretization of each layer (across the structure). This reduces the computation time by an order of magnitude.

\section{\label{sec:discussion}	Numerical and Experimental Results}

\subsection{\label{sec:linear}Linearly polarized waves: Asymmetric transmission}

\begin{figure}[t!]
\centering
\includegraphics[width=1\linewidth]{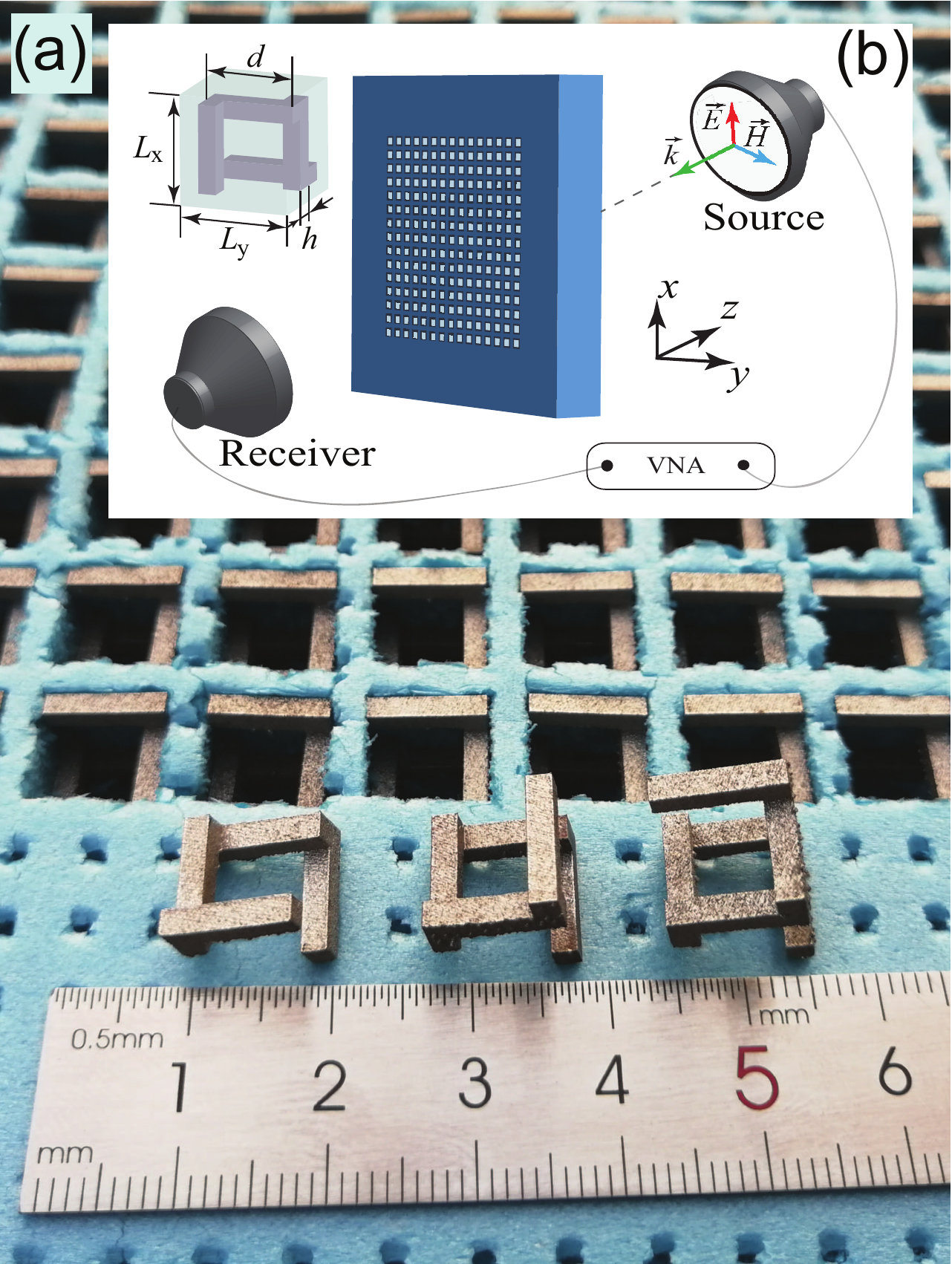}
\caption{(a) A sample of the chiral metasurface and 3D helical particles composed of 4, 5, and 6 rectangular bars produced by direct 3D printing in metal (cobalt--chromium alloy). (b) Sketch of the unit cell and schematic of the experiment.}
\label{fig:sample}
\end{figure}

Consider a chiral metasurface [Fig.~\ref{fig:sample}(a)] operating in the microwave frequency range (1--18~GHz). The metasurface is composed of metallic particles in the form of a stack of metallic bars (dipoles). The bars have rectangular cross-sections and are combined into square helices. The bars are connected to form 4-, 5-, and 6-bar particles (see the inset in the upper-left corner of Fig.~\ref{fig:sample}(b), which shows a sketch of a 4-bar particle). The rectangular bars have thickness $h=2.1$~mm, which is much less than their length $d=10.1$~mm. The size of the unit cell of the metasurface is given by $L_x = L_y = 13.2$~mm and satisfies the subwavelength condition $L/\lambda<0.8$, where $\lambda$ is the wavelength of the illuminating wave. The metallic particles are embedded in a host medium (substrate) with permittivity and permeability close to those of free space. Therefore, the presence of the substrate can be neglected. 

\begin{figure}[t!]
\includegraphics[width=1\linewidth]{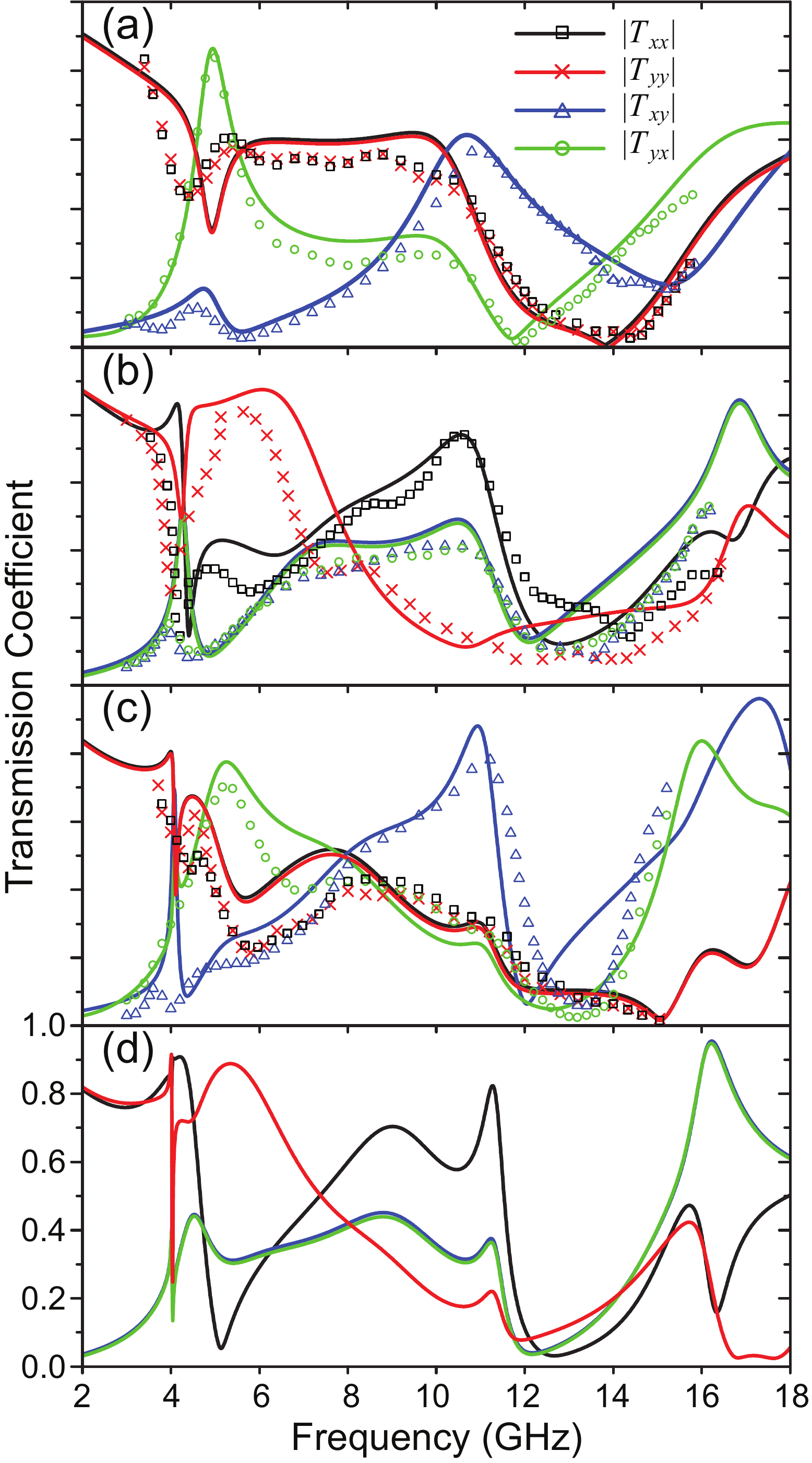}
\caption{Simulated (lines) and measured (points) transmission coefficients of the linearly polarized waves for metasurfaces composed of (a) 4-bar (b) 5-bar, (c) 6-bar, and (d) 7-bar particles.}
\label{fig:l_transmission}
\end{figure}

In the following section, our aim is to investigate the transmission of the normally incident ($\alpha = 0$) $x$-polarized ($\varphi = 0$) and $y$-polarized ($\varphi=\pi/2$) waves through the metasurface with respect to number of bars in the helical particles. The transmission of the $x$-polarized ($y$-polarized) wave can be described by the co-polarization $T_{xx}$  ($T_{yy}$) and cross-polarization $T_{yx}$ ($T_{xy}$) coefficients. These coefficients are elements of the Jones matrix $\mathbf{T}_\textrm{lin}$, which relates the far-fields of the incident (in) and transmitted (out) waves as follows:
\begin{equation}
\begin{pmatrix}
E_x^{\textrm{out}} \\
E_y^{\textrm{out}} 
\end{pmatrix} =
\mathbf{T}_\textrm{lin}
\begin{pmatrix}
E_x^{\textrm{in}}\\
E_y^{\textrm{in}} 
\end{pmatrix} =
\begin{pmatrix}
T_{xx} & T_{xy} \\
 T_{yx}& T_{yy}
\end{pmatrix}
\begin{pmatrix}
E_x^{\textrm{in}}\\
E_y^{\textrm{in}} 
\end{pmatrix}. 
\label{eq:lintr}
\end{equation}
The frequency-dependent transmission coefficients for the metasurfaces composed of helices with different numbers of bars are collected in Fig.~\ref{fig:l_transmission}.

First, we investigate the characteristics of metasurfaces composed of particles with even numbers (4 and 6) of bars. In this case, the simulated data for the co-polarization and cross-polarization transmission coefficients are shown in Figs.~\ref{fig:l_transmission}(a) and \ref{fig:l_transmission}(c). As shown in these figures, the co-polarization transmissions $T_{xx}$ and $T_{yy}$ of both the $x$- and $y$-polarized waves are equal in magnitude because the helical particles with even numbers of bars are symmetric with respect to the orthogonal linearly polarized waves normally incident on opposite sides of the structure (see also the results in Ref.~\onlinecite{Wu_OL_2019} for the  distribution of polarization currents flowing along the bars at the resonant frequencies). In contrast to $T_{xx}$ and $T_{yy}$, the cross-polarization transmission coefficients $T_{yx}$ and $T_{xy}$ of the $x$- and $y$-polarized waves are dissimilar. In particular, these coefficients have distinct spectral positions of peak values, which correspond to the most efficient cross-polarization conversion of the incident wave. The efficiency of such a conversion depends on the number of bars in the helix and can exceed $80\%$ for both $x$- and $y$-polarized waves. The most essential difference between the chiral metasurfaces composed of 4- and 6-bar helices lies in the fact that the transmitted spectrum of the latter structure contains additional low-frequency resonance in the vicinity of 4~GHz (see Fig.~\ref{fig:l_transmission}). 

Since $T_{xy}\neq T_{yx}$, the chiral metasurface composed of helical particles with an even number of rectangular bars results in the asymmetric transmission of the orthogonal linearly polarized waves. Such a transmission of $x$- and $y$-polarized waves is described by the parameters $\Delta^x$ and $\Delta^y$, which show the total intensity difference for the waves propagating through a metasurface in two opposite directions. These parameters can be expressed in terms of two cross-polarization transmission coefficients of $x$- and $y$-polarized waves propagated in one direction:\cite{Menzel_PhysRevLett_2010}
\begin{equation}
\Delta^x = |T_{xy}|^2 - |T_{yx}|^2 = -\Delta^y.
\label{eq:asymtrans}   
\end{equation} 

\begin{figure}[htp]
\includegraphics[width=1\linewidth]{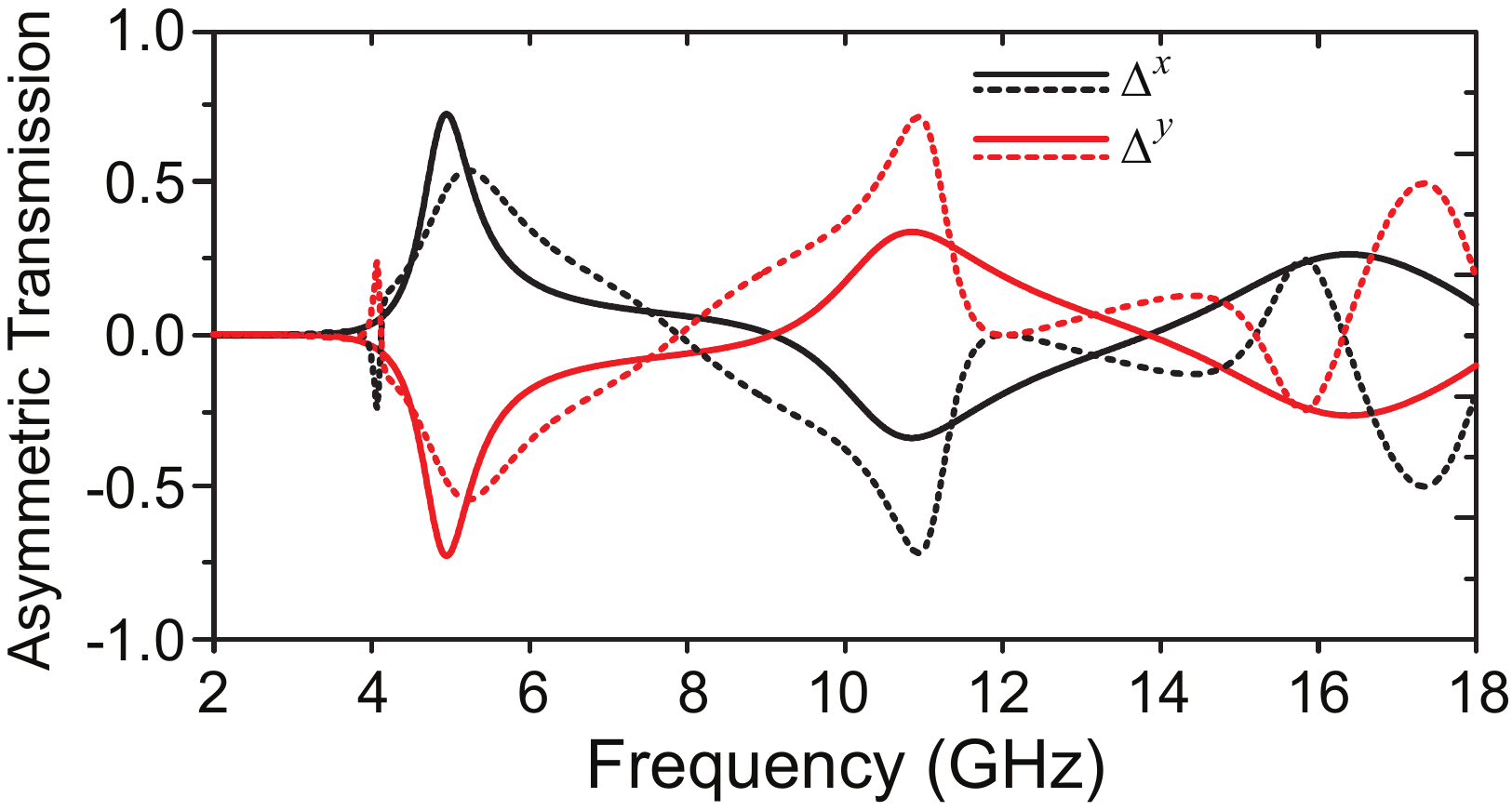}
\caption{Simulated asymmetric transmission of the linearly polarized waves for the metasurfaces composed of 4-bar particles (solid lines) and 6-bar particles (dashed lines).}
\label{fig:l_asymmetric}
\end{figure}

For the metasurfaces made of 4- and 6-bar helical particles, these parameters are shown in Fig.~\ref{fig:l_asymmetric} as functions of wave frequency. One can see that these chiral metasurfaces demonstrate dual-band asymmetric transmissions. As discussed in Ref. \onlinecite{Wu_OL_2019}, this effect is associated with the two lowest resonant modes excited by $x$- and $y$-polarized waves. Each of these modes is characterized by a specific surface current distribution along the rectangular bars, such that the overall current flow is nearly orthogonal to the electric field vector of the incident wave. This current flow gives rise to a transmitted wave with a dominant cross-polarized field component and leads to the maximum cross-polarization transmission coefficient. Since the cross-polarization transmission coefficients of $x$- and $y$-polarized waves attain maxima at two distinct resonance frequencies, one observes the dual-band asymmetric transmission of these waves.

For both metasurfaces, the first and second frequency bands of the asymmetric transmission are located near frequencies of approximately 5  and 11~GHz and correspond to the dominant cross-polarization conversion of the $x$- and $y$-polarized incident waves, respectively [see Figs.~\ref{fig:l_transmission}(a) and \ref{fig:l_transmission}(c)]. In these bands, the cross-polarization transmission of waves of another (orthogonal) polarization is suppressed. For the metasurface with 4-bar square helices, the absolute peak value of the asymmetric transmission of approximately $0.75$ is attained in a low-frequency band. In contrast, for the metasurface made of 6-bar particles, nearly the same peak value of asymmetric transmission is attained at high frequency. Thus, among these metasurfaces, the first one is beneficial for the cross-polarization transmission of low-frequency $x$-polarized waves, while the second one serves as an efficient cross-polarization converter (twist polarizer) of high-frequency $y$-polarized waves.       

A completely different situation occurs in the chiral metasurfaces composed of helical particles with odd numbers of rectangular bars. Figures~\ref{fig:l_transmission}(b) and \ref{fig:l_transmission}(d) show the transmittance for metasurfaces with 5-bar and 7-bar particles, respectively. In this case, contrary to the results depicted in Figs.~\ref{fig:l_transmission}(a) and \ref{fig:l_transmission}(c), the co-polarization transmission coefficients are different for the $x$- and $y$-polarized incident waves. The most intriguing property of the chiral metasurfaces composed of particles with odd numbers of bars is their ability to provide equal (in absolute value) cross-polarization transmission coefficients $T_{xy}$ and $T_{yx}$ throughout the frequency band (i.e., transmission of the $x$- and $y$-polarized waves is symmetrical and satisfies the condition $\Delta^x=\Delta^y=0$). This means that these metasurfaces ensure the same efficiency of cross-polarization conversion for both the $x$- and $y$-polarized incident waves, regardless of their frequency. 
	
In Figs.~\ref{fig:l_transmission}(b) and \ref{fig:l_transmission}(d), one can distinguish a frequency region of broadband cross-polarization conversion with a transmission coefficient of approximately $0.4$. This region expands with increasing number of bars in the helical particles. Compared to the results obtained for the chiral metasurfaces with 4-bar and 6-bar particles, the peak value of the cross-polarization transmission is now attained at a much higher frequency. In the simulation, this frequency has the highest value of $16.85$~GHz for the metasurface made of 5-bar particles and decreases slightly for that with 7-bar particles. This frequency downshift is accompanied by an increase in the maximum efficiency of cross-polarization conversion of the incident wave. For the chiral metasurface with 7-bar helical particles, this efficiency reaches $90\%$. It should be emphasized that the peak efficiency of cross-polarization conversion of the $x$- and $y$-polarized waves is achieved at the same frequency. Therefore, the proposed cross-polarization converter features high performance for any orientation of linearly polarized incident wave. At lower frequencies, a similar result was recently reported for a converter based on a metasurface with a unit cell made of four cylindrical helices.\cite{Faniayeu_APL_2017} Compared to this chiral metasurface, our structure has a simpler design. Therefore, it can be more easily downscaled with the aim of increasing the desired operation frequency of the cross-polarization converter.

To validate the above-described transmission properties of the chiral metasurfaces formed by periodic arrays of square helices, a number of experiments have been performed. 3D printing technology was applied to fabricate the particles from a commercially available cobalt--chromium alloy with a high melting point of 1330$^\circ$C. The 3D-printed helices were fabricated using a Sisma MySint 100 Selective Laser Melting system with a precision of 20--40 \textmu m for use in microwave experiments [see Fig.~\ref{fig:sample}(a)]. The metallic particles were arranged into an array of rectangular holes, which were milled in a custom holder made of a styrofoam material with permittivity $\varepsilon_s = 1.05$. The thickness of holder was $h_s=20.0$~mm. The experimental prototype chiral metasurface consisted of $15 \times 15$ unit cells, corresponding to 225 helical particles. The dimensions of the particles and unit cells were identical to those used in the simulations. The lattice of particles, which were arranged into the custom holder, effectively formed the metasurface prototype. 

The metasurface prototype was placed in an echoic chamber and illuminated by normally incident $x$-polarized (or $y$-polarized) waves radiated and received by a pair of horn antennas (HD-10180DRHA10SZJ) with an operating range of 1--18~GHz (in fact, we restricted the measurements to the bandwidth of 3--15~GHz to exclude the non-optimal characteristics of the antennas). The antennas were connected to the ports of a Keysight E5071C Vector Network Analyzer (VNA) by 50-\textOmega{} coaxial cables [a schematic of the experimental setup is shown in Fig.~\ref{fig:sample}(b)]. 

The transmissions of linearly $x$- and $y$-polarized waves were measured for the chiral metasurfaces made of 4-bar, 5-bar, and 6-bar square helices. The measurement results are shown in Fig.~\ref{fig:l_transmission} as scattered data points. Reasonable agreement can be observed between the simulation and measurement results, although some discrepancies resulted from the imperfections inherent in the experiment. Thus, the experimental data support our theoretical findings concerning the effect of the number of bars in the helical particles on the metasurface transmittance. 

\subsection{\label{sec:parameters}Effects of bar thickness and length}

\begin{figure} [t!]
\includegraphics[width=1\linewidth]{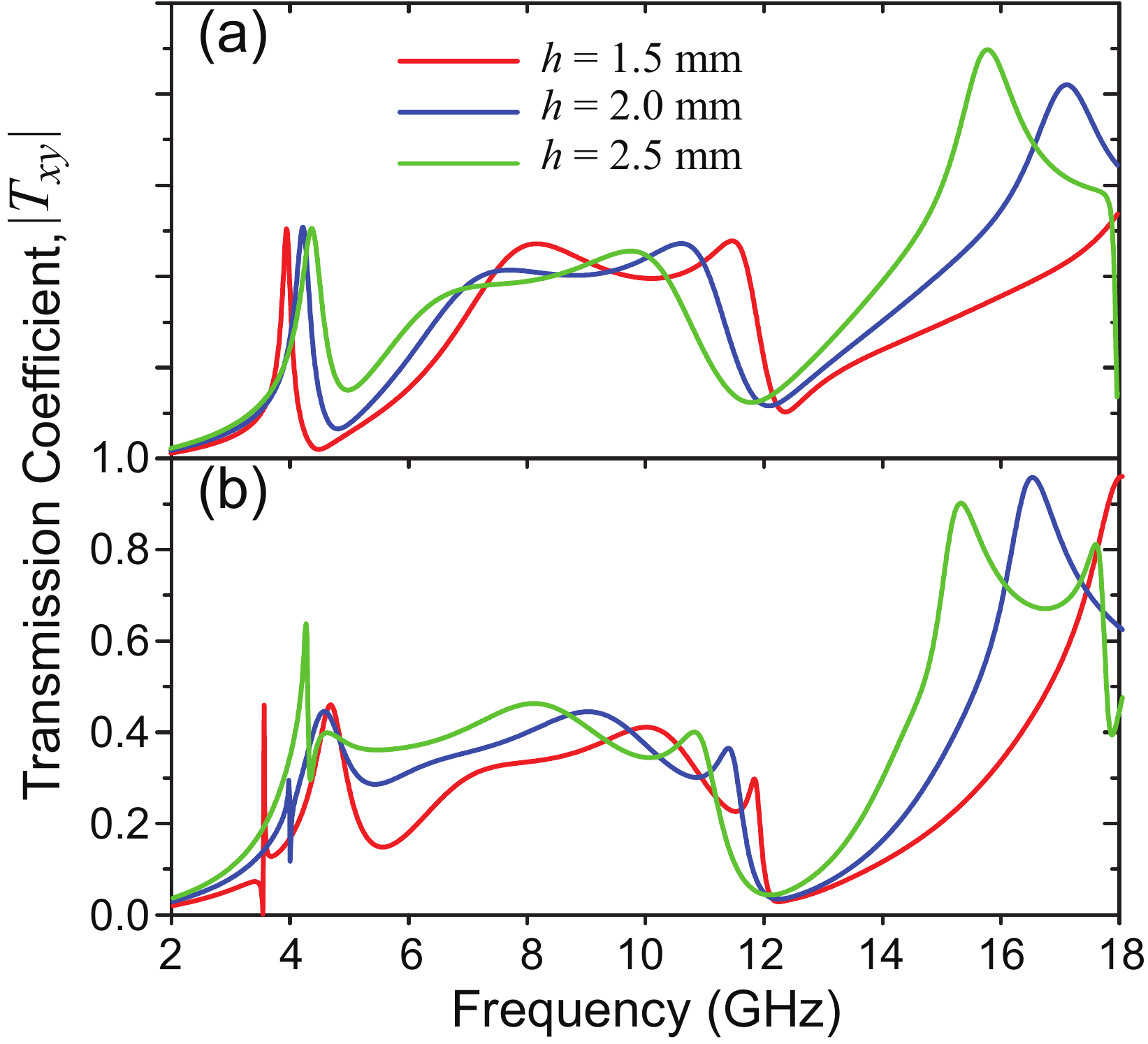}
\caption{Simulated cross-polarization transmission of linearly polarized waves as a function of bar thickness for the metasurfaces composed of (a) 5-bar particles and (b) 7-bar particles.}
\label{fig:thickness}
\end{figure}

\begin{figure}[t!]
\includegraphics[width=1\linewidth]{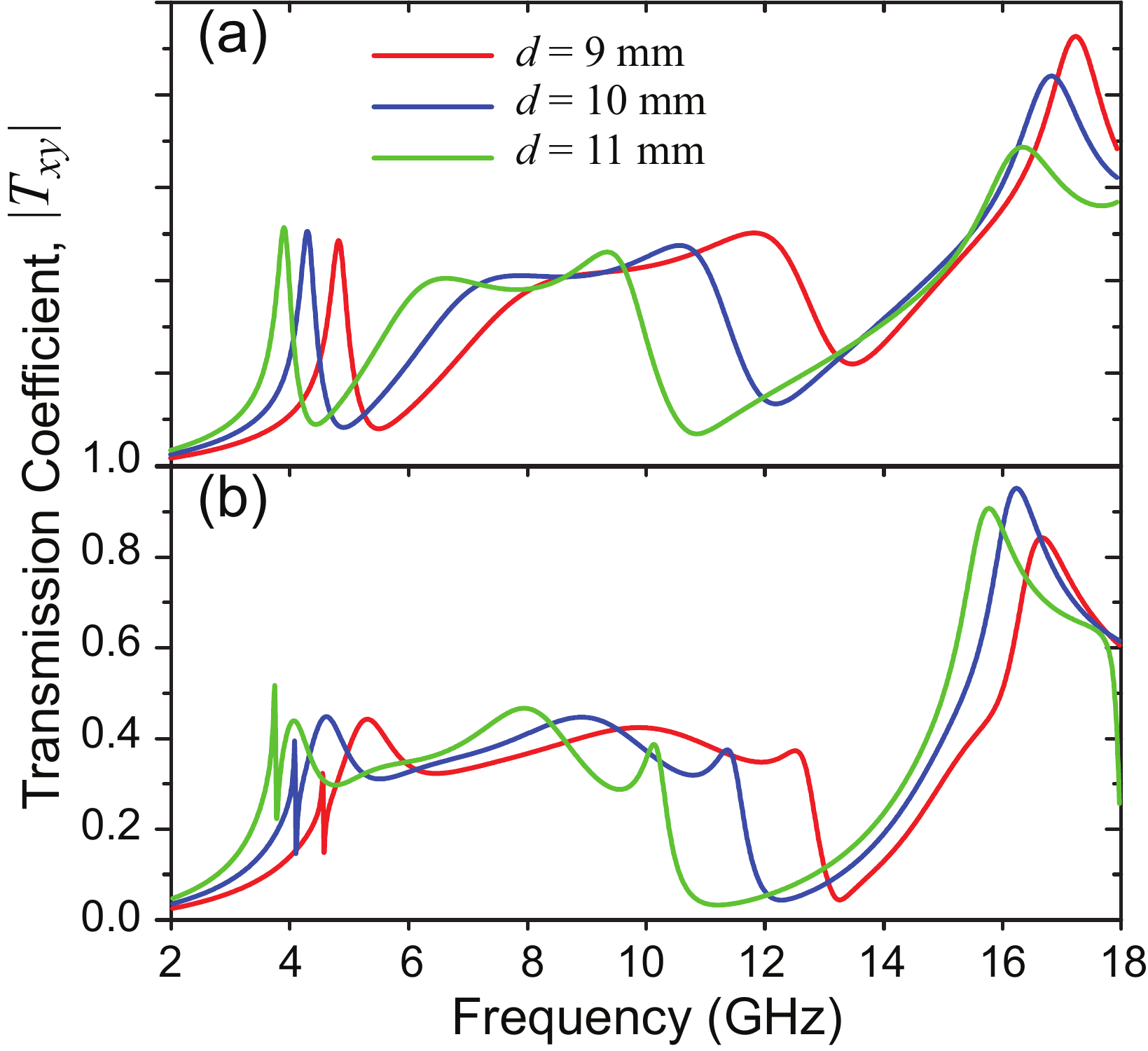}
\caption{Same as Fig.~\ref{fig:thickness} but for different bar lenghts.}
\label{fig:length}
\end{figure}

Next, we theoretically investigate the transmittance of a proposed metasurface as a function of the structural parameters of the rectangular helical particles. Such an investigation is useful in determining the sensitivity of the cross-polarization conversion to the fabrication tolerance of the helical particles. In this investigation, we consider metasurfaces composed of 5-bar and 7-bar square helices, which provide transmission of linearly polarized waves with equal absolute values of cross-polarization coefficients (i.e., $|T_{xy}|=|T_{yx}|$) for any wave frequency. These coefficients are studied with respect to the thickness $h$ and length $d$ of the rectangular bars involved in the particle design. The results are summarized in Figs.~\ref{fig:thickness} and \ref{fig:length}. 

From Fig.~\ref{fig:thickness}, we note that the transmission characteristics of the metasurfaces are highly sensitive to the bar thickness. Increasing the bar thickness by 0.5~mm induces a frequency downshift of high-transmission resonance of up to 1~GHz. In this process, the peak value of cross-polarized transmission can either increase or decrease depending on the number of bars in the particles. For lower frequencies, the above-discussed cross-polarization conversion, which has a broad frequency band, is also distinctly affected by $h$. As $h$ increases, the cross-polarization transmission coefficients becomes more uniformly distributed within this frequency band, which shifts toward lower frequencies. 

Compared to the bar thickness $h$, the bar length $d$ has a more regular effect on cross-polarized transmission (Fig.~\ref{fig:length}). As $d$ decreases, the transmission resonances shift to higher frequencies, while the frequency band of broadband cross-polarization conversion expands. Interestingly, in the chosen frequency band, the length $d=10$~mm provides the highest peak value of cross-polarized transmission for the metasurface composed of 7-bar particles. In contrast, for the metasurface with 5-bar particles, this value increases with decreasing bar length. Thus, the geometric parameters of the bars can be optimized to achieve highly efficient cross-polarization conversion of the linearly polarized waves at a desired frequency.

\subsection{\label{sec:circular}Circularly polarized waves: Circular dichroism}

A unique feature of metasurfaces composed of true 3D chiral particles is circular dichroism; that is, these metasurfaces are characterized by different transmittances for the right-handed (RCP) and left-handed (LCP) circularly polarized waves.\cite{berova_book_2000} The transmission coefficients of the RCP and LCP waves can be obtained directly from those of the linearly polarized waves as follows:
\begin{equation}
\begin{split}
\mathbf{T}_\textrm{circ} &= \begin{pmatrix}
T_{++} & T_{+-} \\
T_{-+} & T_{--}
\end{pmatrix} = \Lambda^{-1}\mathbf{T}_\textrm{lin}\Lambda \\
&=\frac{1}{2}
\begin{pmatrix}
T_{xx} + T_{yy} & T_{xx} - T_{yy} \\
T_{xx} - T_{yy} & T_{xx} + T_{yy}
\end{pmatrix} \\
&+\frac{i}{2}
\begin{pmatrix}
T_{xy} - T_{yx} & -T_{xy} - T_{yx} \\
T_{xy} + T_{yx} & -T_{yx} + T_{xy}
\end{pmatrix},
\end{split}  
\label{eq:rcplcp}
\end{equation}
where $\Lambda=\frac{1}{\sqrt{2}}\begin{pmatrix} 1 & 1 \\ i & -i\end{pmatrix}$ is the change-of-basis matrix. The lower signs ``$+$'' and ``$-$'' are used to describe the RCP and LCP waves, respectively, and $T_{++}$ ($T_{--}$) and $T_{+-}$ ($T_{-+}$) are their co-polarization and cross-polarization transmission coefficients. 

\begin{figure}[t!]
\includegraphics[width=0.98\linewidth]{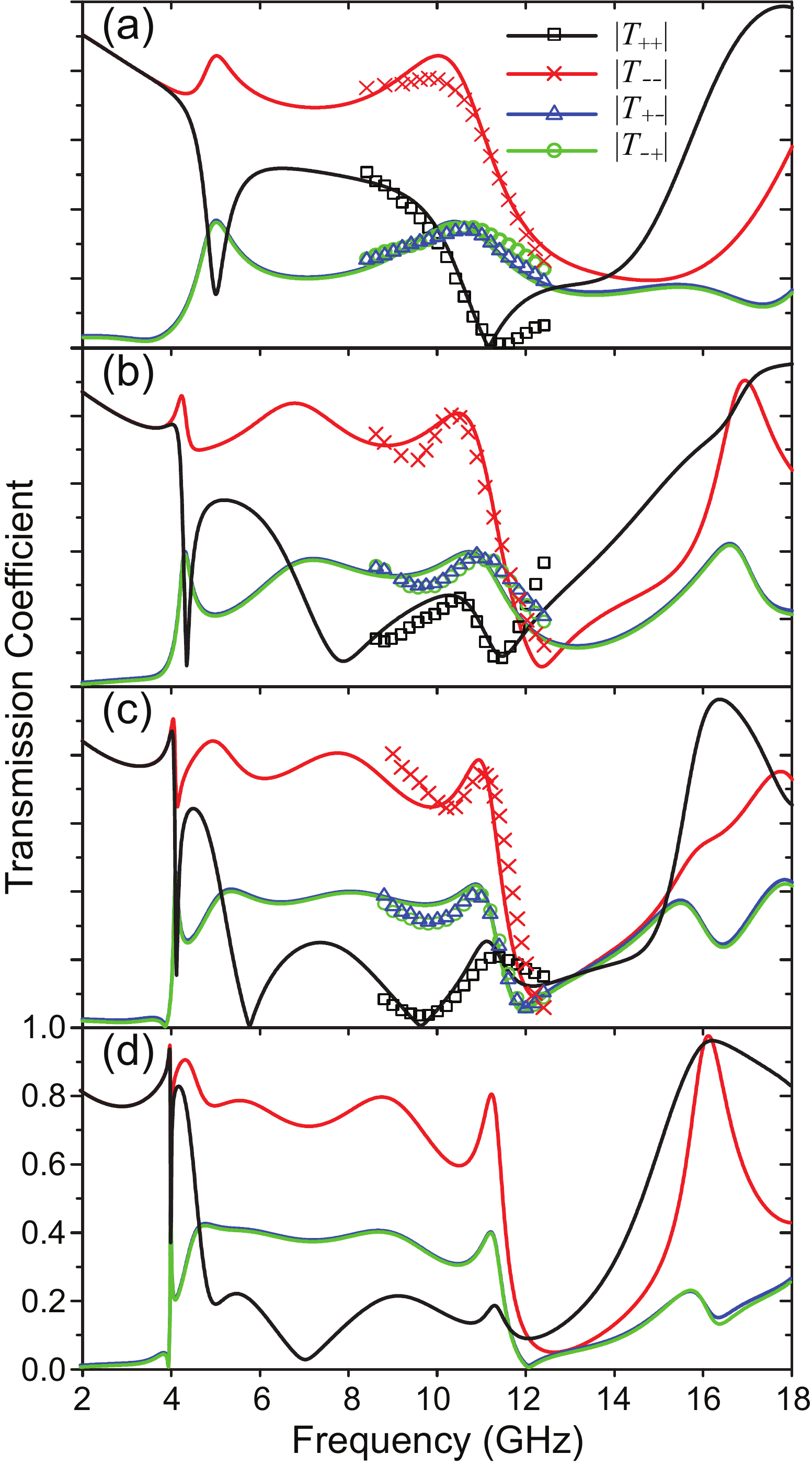}
\caption{Same as Fig.~\ref{fig:l_transmission} but for circularly polarized waves.}
\label{fig:c_transmission}
\end{figure}

Figure~\ref{fig:c_transmission} shows the simulated transmission coefficients of the RCP and LCP waves for chiral metasurfaces made of particles with different numbers of bars (from 4 to 7). One can notice that the condition $|T_{+-}|=|T_{-+}|$ holds true for both even and odd numbers of bars. Thus, these metasurfaces provide symmetric transmission of RCP and LCP waves.

\begin{figure}[t!]
\includegraphics[width=0.98\linewidth]{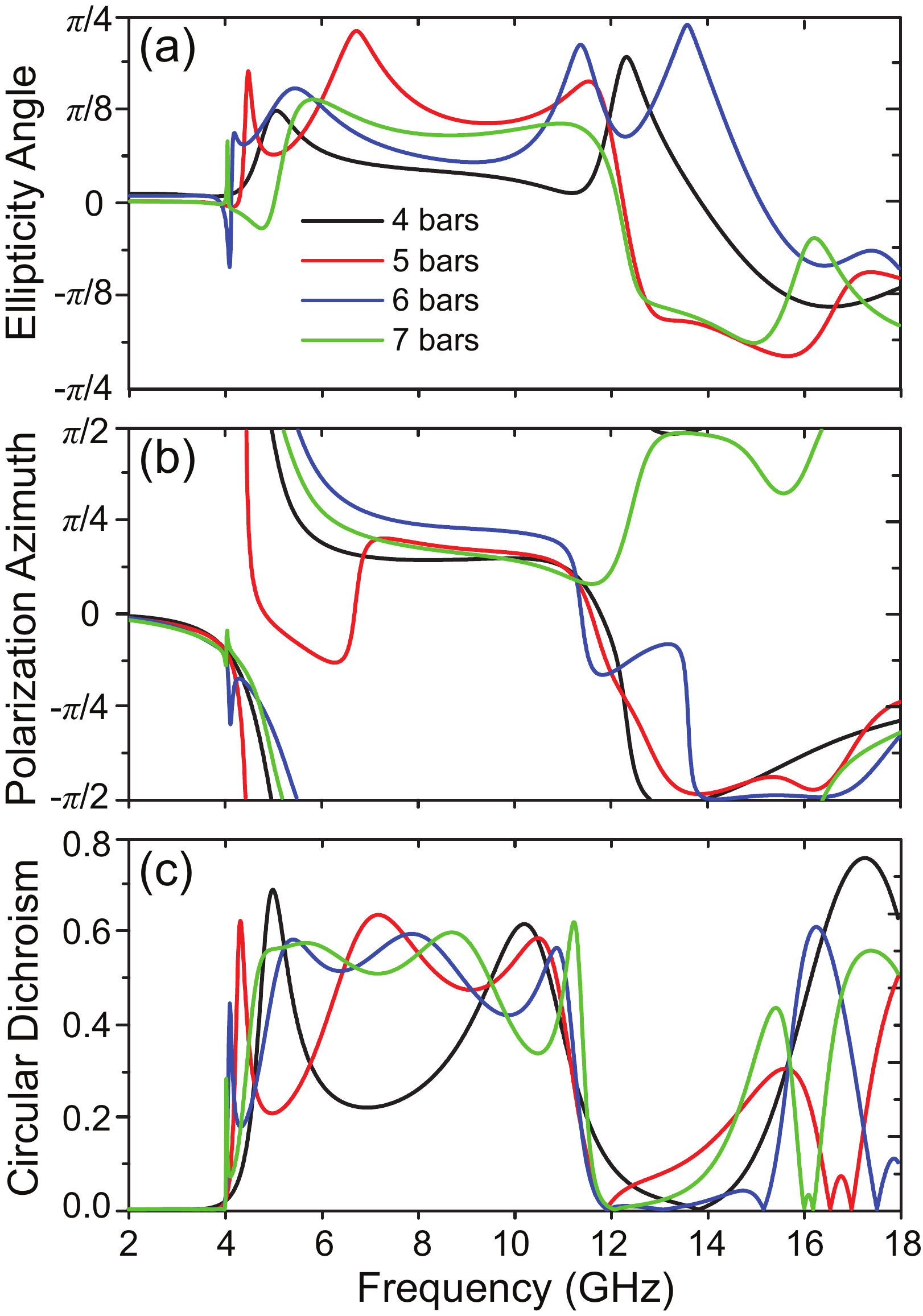}
\caption{(a) Ellipticity angle $\eta$, (b) polarization azimuth $\theta$ (both in radians), and (c) broadband circular dichroism of transmitted waves for metasurfaces composed of different sets of particles.}
\label{fig:ellipticity}
\end{figure}

Using the co-polarization and cross-polarization transmission coefficients, one can determine the polarization state of the transmitted field from the following well-known formulae: \cite{Collett_1993} $\tan 2\theta = U_2/U_1$ and $\sin 2\eta = U_3/U_0$,  where $\theta$ is the polarization azimuth, $\eta$ is the ellipticity angle, and $U_j$ ($j = 0,\ldots,3$) are the Stokes parameters expressed in terms of the electric field components in the right-handed orthogonal coordinate system. According to the definition of the Stokes parameters, the ellipticity equals zero, $-\pi/4$, and $+\pi/4$ for linearly polarized, LCP, and RCP waves, respectively. In the case of $0<|\eta|<\pi/4$, the transmitted field is elliptically polarized. For the field transmitted through the chiral metasurface, the ellipticity and polarization azimuth are shown as functions of frequency in Figs.~\ref{fig:ellipticity}(a) and \ref{fig:ellipticity}(b), respectively. These figures show that this metasurface generally transforms a linearly polarized wave into an elliptically polarized wave in the chosen frequency band. The exceptions are two resonant frequencies of approximately 7 and 14~GHz for the metasurfaces composed of 5-bar and 6-bar particles, respectively. For these frequencies, the transmitted field exhibits nearly circular polarization.

Co-polarization transmission is generally different for orthogonally polarized circular waves, leading to non-zero circular dichroism: $\textrm{CD}=|T_{++}|^2- |T_{--}|^2$. As shown in Fig.~\ref{fig:ellipticity}(c), the circular dichroism has a narrowband resonant nature at high frequencies, while it oscillates near approximately 0.6 in the broad frequency band between 4 and 12~GHz. Within this band, the largest value of circular dichroism (in excess of 0.7) is attained for the metasurface with 4-bar helical particles. However, this metasurface exhibits large variation in circular dichroism with frequency. The distribution of circular dichroism along the frequency band becomes more uniform as the number of bars in the helical particles increases. During this process, the frequency bandwidth of the strong circular dichroism increases slightly. For the chiral metasurface composed of 7-bar square helices, this bandwidth exceeds 90\% with respect to the central frequency of 7.6~GHz. This demonstrates the ability of this metasurface to provide strong circular dichroism in a broad frequency band.

Our simulations are supported by experimental data obtained from the direct measurements of the transmission of circularly polarized waves through the metasurfaces made of 4-bar, 5-bar, and 6-bar particles. In the experiments, we used HD-80124CPHA16SR(L) antennas to radiate and receive RCP (LCP) waves in a relatively narrow frequency band between 8 and 12.4~GHz. For these frequencies, the measurement results match well with the simulations, thereby confirming the transmission characteristics of the chiral metasurface under incident RCP and LCP waves. The agreement between theory and experiment also provides evidence supporting the theoretical approach we used to analyze multilayered metasurfaces formed by periodic arrays of metallic square bars.

\section{\label{sec:level1}Conclusions}

We have presented a detailed analysis of the transmission of linearly and circularly polarized waves incident on chiral metasurfaces formed from arrays of metallic helical particles in the form of square helices. The analysis involved both theoretical and experimental investigations. In the theoretical investigations, the rigorous approach of equivalent polarization currents was extended to simulate wave scattering from the multilayered structure. This approach was validated against experimental measurements. The transmission characteristics of linearly polarized waves were found to be essentially different for metasurfaces made of square helices with even and odd numbers of metallic bars. In the case of chiral metasurfaces with even numbers of bars of helical particles, the $x$- and $y$-linearly polarized waves feature the same co-polarization transmission, while their cross-polarization transmission coefficients are different and reach peak values at distinct frequencies. As a result, the metasurface results in the dual-band asymmetric transmission of these waves. In contrast, for metasurfaces with odd numbers of bars of square helices, the co-polarization transmission coefficients of $x$- and $y$-linearly polarized waves differ in value; however, the cross-polarization conversion of these waves is characterized by the same efficiency, regardless of the frequency of the incident wave. This efficiency was found to reach 90\% for the chiral metasurface with 7-bar helical particles. Thus, this metasurface behaves like a high-performance cross-polarization converter that is independent of the orientation of the linearly polarized incident waves. 

The findings show that the dimensions of metallic bars incorporated into square helices can be optimized to achieve highly efficient cross-polarization conversion at the desired wave frequency. In addition, for circularly polarized incident waves, the metasurface composed of square helices was shown to exhibit broadband circular dichroism with a maximum value of 0.7. The relative frequency bandwidth of circular dichroism was shown to exceed 90\%. The obtained results demonstrate that particles in the form of square helices provide broad variability in the transmission properties of chiral metasurfaces. Therefore, such chiral metasurfaces are expected to find wide application in efficient polarization-control and polarization-transmission devices. 

\section*{\label{sec:ack}Acknowledgments}
V.V.Y and V.R.T. thank Prof. S. L. Prosvirnin for long-term cooperation, fruitful discussions, and suggestions.

\bibliography{chiral_full}

\begin{thebibliography}{50}%
\makeatletter
\providecommand \@ifxundefined [1]{%
 \@ifx{#1\undefined}
}%
\providecommand \@ifnum [1]{%
 \ifnum #1\expandafter \@firstoftwo
 \else \expandafter \@secondoftwo
 \fi
}%
\providecommand \@ifx [1]{%
 \ifx #1\expandafter \@firstoftwo
 \else \expandafter \@secondoftwo
 \fi
}%
\providecommand \natexlab [1]{#1}%
\providecommand \enquote  [1]{``#1''}%
\providecommand \bibnamefont  [1]{#1}%
\providecommand \bibfnamefont [1]{#1}%
\providecommand \citenamefont [1]{#1}%
\providecommand \href@noop [0]{\@secondoftwo}%
\providecommand \href [0]{\begingroup \@sanitize@url \@href}%
\providecommand \@href[1]{\@@startlink{#1}\@@href}%
\providecommand \@@href[1]{\endgroup#1\@@endlink}%
\providecommand \@sanitize@url [0]{\catcode `\\12\catcode `\$12\catcode
  `\&12\catcode `\#12\catcode `\^12\catcode `\_12\catcode `\%12\relax}%
\providecommand \@@startlink[1]{}%
\providecommand \@@endlink[0]{}%
\providecommand \url  [0]{\begingroup\@sanitize@url \@url }%
\providecommand \@url [1]{\endgroup\@href {#1}{\urlprefix }}%
\providecommand \urlprefix  [0]{URL }%
\providecommand \Eprint [0]{\href }%
\providecommand \doibase [0]{http://dx.doi.org/}%
\providecommand \selectlanguage [0]{\@gobble}%
\providecommand \bibinfo  [0]{\@secondoftwo}%
\providecommand \bibfield  [0]{\@secondoftwo}%
\providecommand \translation [1]{[#1]}%
\providecommand \BibitemOpen [0]{}%
\providecommand \bibitemStop [0]{}%
\providecommand \bibitemNoStop [0]{.\EOS\space}%
\providecommand \EOS [0]{\spacefactor3000\relax}%
\providecommand \BibitemShut  [1]{\csname bibitem#1\endcsname}%
\let\auto@bib@innerbib\@empty
\bibitem [{\citenamefont {Barron}(2004)}]{Barron_book_2004}%
  \BibitemOpen
  \bibfield  {author} {\bibinfo {author} {\bibfnamefont {L.~D.}\ \bibnamefont
  {Barron}},\ }\href@noop {} {\emph {\bibinfo {title} {Molecular Light
  Scattering and Optical Activity}}},\ \bibinfo {edition} {2nd}\ ed.\ (\bibinfo
   {publisher} {Cambridge University Press, Cambridge},\ \bibinfo {year}
  {2004})\BibitemShut {NoStop}%
\bibitem [{\citenamefont {Hecht}(2016)}]{Hecht_2016}%
  \BibitemOpen
  \bibfield  {author} {\bibinfo {author} {\bibfnamefont {E.}~\bibnamefont
  {Hecht}},\ }\href@noop {} {\emph {\bibinfo {title} {Optics}}},\ \bibinfo
  {edition} {5th}\ ed.\ (\bibinfo  {publisher} {Pearson Education, Harlow},\
  \bibinfo {year} {2016})\BibitemShut {NoStop}%
\bibitem [{\citenamefont {Vallius}\ \emph {et~al.}(2003)\citenamefont
  {Vallius}, \citenamefont {Jefimovs}, \citenamefont {Turunen}, \citenamefont
  {Vahimaa},\ and\ \citenamefont {Svirko}}]{Svirko_ApplPhysLett_2003}%
  \BibitemOpen
  \bibfield  {author} {\bibinfo {author} {\bibfnamefont {T.}~\bibnamefont
  {Vallius}}, \bibinfo {author} {\bibfnamefont {K.}~\bibnamefont {Jefimovs}},
  \bibinfo {author} {\bibfnamefont {J.}~\bibnamefont {Turunen}}, \bibinfo
  {author} {\bibfnamefont {P.}~\bibnamefont {Vahimaa}}, \ and\ \bibinfo
  {author} {\bibfnamefont {Y.}~\bibnamefont {Svirko}},\ }\bibfield  {title}
  {\enquote {\bibinfo {title} {Optical activity in subwavelength-period arrays
  of chiral metallic particles},}\ }\href {\doibase 10.1063/1.1592015}
  {\bibfield  {journal} {\bibinfo  {journal} {Appl. Phys. Lett.}\ }\textbf
  {\bibinfo {volume} {83}},\ \bibinfo {pages} {234--236} (\bibinfo {year}
  {2003})}\BibitemShut {NoStop}%
\bibitem [{\citenamefont {Prosvirnin}\ and\ \citenamefont
  {Zheludev}(2005)}]{Prosvirnin_PhysRevE_2005}%
  \BibitemOpen
  \bibfield  {author} {\bibinfo {author} {\bibfnamefont {S.~L.}\ \bibnamefont
  {Prosvirnin}}\ and\ \bibinfo {author} {\bibfnamefont {N.~I.}\ \bibnamefont
  {Zheludev}},\ }\bibfield  {title} {\enquote {\bibinfo {title} {Polarization
  effects in the diffraction of light by a planar chiral structure},}\ }\href
  {\doibase 10.1103/PhysRevE.71.037603} {\bibfield  {journal} {\bibinfo
  {journal} {Phys. Rev. E}\ }\textbf {\bibinfo {volume} {71}},\ \bibinfo
  {pages} {037603} (\bibinfo {year} {2005})}\BibitemShut {NoStop}%
\bibitem [{\citenamefont {Fedotov}\ \emph {et~al.}(2006)\citenamefont
  {Fedotov}, \citenamefont {Mladyonov}, \citenamefont {Prosvirnin},
  \citenamefont {Rogacheva}, \citenamefont {Chen},\ and\ \citenamefont
  {Zheludev}}]{Fedotov_PhysRevLett_2006}%
  \BibitemOpen
  \bibfield  {author} {\bibinfo {author} {\bibfnamefont {V.~A.}\ \bibnamefont
  {Fedotov}}, \bibinfo {author} {\bibfnamefont {P.~L.}\ \bibnamefont
  {Mladyonov}}, \bibinfo {author} {\bibfnamefont {S.~L.}\ \bibnamefont
  {Prosvirnin}}, \bibinfo {author} {\bibfnamefont {A.~V.}\ \bibnamefont
  {Rogacheva}}, \bibinfo {author} {\bibfnamefont {Y.}~\bibnamefont {Chen}}, \
  and\ \bibinfo {author} {\bibfnamefont {N.~I.}\ \bibnamefont {Zheludev}},\
  }\bibfield  {title} {\enquote {\bibinfo {title} {Asymmetric propagation of
  electromagnetic waves through a planar chiral structure},}\ }\href {\doibase
  10.1103/PhysRevLett.97.167401} {\bibfield  {journal} {\bibinfo  {journal}
  {Phys. Rev. Lett.}\ }\textbf {\bibinfo {volume} {97}},\ \bibinfo {pages}
  {167401} (\bibinfo {year} {2006})}\BibitemShut {NoStop}%
\bibitem [{\citenamefont {Prosvirnin}\ and\ \citenamefont
  {Zheludev}(2009)}]{Prosvirnin_JOpt_2009}%
  \BibitemOpen
  \bibfield  {author} {\bibinfo {author} {\bibfnamefont {S.~L.}\ \bibnamefont
  {Prosvirnin}}\ and\ \bibinfo {author} {\bibfnamefont {N.~I.}\ \bibnamefont
  {Zheludev}},\ }\bibfield  {title} {\enquote {\bibinfo {title} {Analysis of
  polarization transformations by a planar chiral array of complex-shaped
  particles},}\ }\href {http://stacks.iop.org/1464-4258/11/i=7/a=074002}
  {\bibfield  {journal} {\bibinfo  {journal} {J. Opt. A: Pure Appl. Opt.}\
  }\textbf {\bibinfo {volume} {11}},\ \bibinfo {pages} {074002} (\bibinfo
  {year} {2009})}\BibitemShut {NoStop}%
\bibitem [{\citenamefont {Plum}\ \emph {et~al.}(2009)\citenamefont {Plum},
  \citenamefont {Zhou}, \citenamefont {Dong}, \citenamefont {Fedotov},
  \citenamefont {Koschny}, \citenamefont {Soukoulis},\ and\ \citenamefont
  {Zheludev}}]{Plum_PhysRevB_2009}%
  \BibitemOpen
  \bibfield  {author} {\bibinfo {author} {\bibfnamefont {E.}~\bibnamefont
  {Plum}}, \bibinfo {author} {\bibfnamefont {J.}~\bibnamefont {Zhou}}, \bibinfo
  {author} {\bibfnamefont {J.}~\bibnamefont {Dong}}, \bibinfo {author}
  {\bibfnamefont {V.~A.}\ \bibnamefont {Fedotov}}, \bibinfo {author}
  {\bibfnamefont {T.}~\bibnamefont {Koschny}}, \bibinfo {author} {\bibfnamefont
  {C.~M.}\ \bibnamefont {Soukoulis}}, \ and\ \bibinfo {author} {\bibfnamefont
  {N.~I.}\ \bibnamefont {Zheludev}},\ }\bibfield  {title} {\enquote {\bibinfo
  {title} {Metamaterial with negative index due to chirality},}\ }\href
  {\doibase 10.1103/PhysRevB.79.035407} {\bibfield  {journal} {\bibinfo
  {journal} {Phys. Rev. B}\ }\textbf {\bibinfo {volume} {79}},\ \bibinfo
  {pages} {035407} (\bibinfo {year} {2009})}\BibitemShut {NoStop}%
\bibitem [{\citenamefont {Plum}, \citenamefont {Fedotov},\ and\ \citenamefont
  {Zheludev}(2010)}]{plum2010asymmetric}%
  \BibitemOpen
  \bibfield  {author} {\bibinfo {author} {\bibfnamefont {E.}~\bibnamefont
  {Plum}}, \bibinfo {author} {\bibfnamefont {V.}~\bibnamefont {Fedotov}}, \
  and\ \bibinfo {author} {\bibfnamefont {N.}~\bibnamefont {Zheludev}},\
  }\bibfield  {title} {\enquote {\bibinfo {title} {Asymmetric transmission: a
  generic property of two-dimensional periodic patterns},}\ }\href
  {http://stacks.iop.org/2040-8986/13/i=2/a=024006} {\bibfield  {journal}
  {\bibinfo  {journal} {J. Opt.}\ }\textbf {\bibinfo {volume} {13}},\ \bibinfo
  {pages} {024006} (\bibinfo {year} {2010})}\BibitemShut {NoStop}%
\bibitem [{\citenamefont {Engelbrecht}\ \emph {et~al.}(2010)\citenamefont
  {Engelbrecht}, \citenamefont {Wunderlich}, \citenamefont {Shuvaev},\ and\
  \citenamefont {Pimenov}}]{Engelbrecht_ApplPhysLett_2010}%
  \BibitemOpen
  \bibfield  {author} {\bibinfo {author} {\bibfnamefont {S.}~\bibnamefont
  {Engelbrecht}}, \bibinfo {author} {\bibfnamefont {M.}~\bibnamefont
  {Wunderlich}}, \bibinfo {author} {\bibfnamefont {A.~M.}\ \bibnamefont
  {Shuvaev}}, \ and\ \bibinfo {author} {\bibfnamefont {A.}~\bibnamefont
  {Pimenov}},\ }\bibfield  {title} {\enquote {\bibinfo {title} {Colossal
  optical activity of split-ring resonator arrays for millimeter waves},}\
  }\href {\doibase 10.1063/1.3481699} {\bibfield  {journal} {\bibinfo
  {journal} {Appl. Phys. Lett.}\ }\textbf {\bibinfo {volume} {97}},\ \bibinfo
  {pages} {081116} (\bibinfo {year} {2010})}\BibitemShut {NoStop}%
\bibitem [{\citenamefont {Shi}\ \emph {et~al.}(2013)\citenamefont {Shi},
  \citenamefont {Liu}, \citenamefont {Yu}, \citenamefont {Lv}, \citenamefont
  {Zhu}, \citenamefont {{Feng Ma}},\ and\ \citenamefont {{Jun
  Cui}}}]{Shi_ApplPhysLett_2013}%
  \BibitemOpen
  \bibfield  {author} {\bibinfo {author} {\bibfnamefont {J.}~\bibnamefont
  {Shi}}, \bibinfo {author} {\bibfnamefont {X.}~\bibnamefont {Liu}}, \bibinfo
  {author} {\bibfnamefont {S.}~\bibnamefont {Yu}}, \bibinfo {author}
  {\bibfnamefont {T.}~\bibnamefont {Lv}}, \bibinfo {author} {\bibfnamefont
  {Z.}~\bibnamefont {Zhu}}, \bibinfo {author} {\bibfnamefont {H.}~\bibnamefont
  {{Feng Ma}}}, \ and\ \bibinfo {author} {\bibfnamefont {T.}~\bibnamefont {{Jun
  Cui}}},\ }\bibfield  {title} {\enquote {\bibinfo {title} {Dual-band
  asymmetric transmission of linear polarization in bilayered chiral
  metamaterial},}\ }\href {\doibase 10.1063/1.4805075} {\bibfield  {journal}
  {\bibinfo  {journal} {Appl. Phys. Lett.}\ }\textbf {\bibinfo {volume}
  {102}},\ \bibinfo {pages} {191905} (\bibinfo {year} {2013})}\BibitemShut
  {NoStop}%
\bibitem [{\citenamefont {Hannam}\ \emph {et~al.}(2013)\citenamefont {Hannam},
  \citenamefont {Powell}, \citenamefont {Shadrivov},\ and\ \citenamefont
  {Kivshar}}]{Shadrivov_ApplPhyLett_2013}%
  \BibitemOpen
  \bibfield  {author} {\bibinfo {author} {\bibfnamefont {K.}~\bibnamefont
  {Hannam}}, \bibinfo {author} {\bibfnamefont {D.~A.}\ \bibnamefont {Powell}},
  \bibinfo {author} {\bibfnamefont {I.~V.}\ \bibnamefont {Shadrivov}}, \ and\
  \bibinfo {author} {\bibfnamefont {Y.~S.}\ \bibnamefont {Kivshar}},\
  }\bibfield  {title} {\enquote {\bibinfo {title} {Dispersionless optical
  activity in metamaterials},}\ }\href {\doibase 10.1063/1.4807438} {\bibfield
  {journal} {\bibinfo  {journal} {Appl. Phys. Lett.}\ }\textbf {\bibinfo
  {volume} {102}},\ \bibinfo {pages} {201121} (\bibinfo {year}
  {2013})}\BibitemShut {NoStop}%
\bibitem [{\citenamefont {Polevoy}\ \emph {et~al.}(2013)\citenamefont
  {Polevoy}, \citenamefont {Prosvirnin}, \citenamefont {Tarapov},\ and\
  \citenamefont {Tuz}}]{Polevoy_2013}%
  \BibitemOpen
  \bibfield  {author} {\bibinfo {author} {\bibfnamefont {S.~Y.}\ \bibnamefont
  {Polevoy}}, \bibinfo {author} {\bibfnamefont {S.~L.}\ \bibnamefont
  {Prosvirnin}}, \bibinfo {author} {\bibfnamefont {S.~I.}\ \bibnamefont
  {Tarapov}}, \ and\ \bibinfo {author} {\bibfnamefont {V.~R.}\ \bibnamefont
  {Tuz}},\ }\bibfield  {title} {\enquote {\bibinfo {title} {Resonant features
  of planar {Faraday} metamaterial with high structural symmetry -- {Study} of
  properties of a 4-fold array of planar chiral rosettes placed on a ferrite
  substrate},}\ }\href {\doibase 10.1051/epjap/2013120320} {\bibfield
  {journal} {\bibinfo  {journal} {Eur. Phys. J. Appl. Phys.}\ }\textbf
  {\bibinfo {volume} {61}},\ \bibinfo {pages} {30501} (\bibinfo {year}
  {2013})}\BibitemShut {NoStop}%
\bibitem [{\citenamefont {Song}\ \emph {et~al.}(2016)\citenamefont {Song},
  \citenamefont {Ding}, \citenamefont {Su}, \citenamefont {Liu}, \citenamefont
  {Luo}, \citenamefont {Zhao}, \citenamefont {Bhattarai},\ and\ \citenamefont
  {Zhou}}]{Song_JApplPhys_2016}%
  \BibitemOpen
  \bibfield  {author} {\bibinfo {author} {\bibfnamefont {K.}~\bibnamefont
  {Song}}, \bibinfo {author} {\bibfnamefont {C.}~\bibnamefont {Ding}}, \bibinfo
  {author} {\bibfnamefont {Z.}~\bibnamefont {Su}}, \bibinfo {author}
  {\bibfnamefont {Y.}~\bibnamefont {Liu}}, \bibinfo {author} {\bibfnamefont
  {C.}~\bibnamefont {Luo}}, \bibinfo {author} {\bibfnamefont {X.}~\bibnamefont
  {Zhao}}, \bibinfo {author} {\bibfnamefont {K.}~\bibnamefont {Bhattarai}}, \
  and\ \bibinfo {author} {\bibfnamefont {J.}~\bibnamefont {Zhou}},\ }\bibfield
  {title} {\enquote {\bibinfo {title} {Planar composite chiral metamaterial
  with broadband dispersionless polarization rotation and high transmission},}\
  }\href {\doibase 10.1063/1.4972977} {\bibfield  {journal} {\bibinfo
  {journal} {J. Appl. Phys.}\ }\textbf {\bibinfo {volume} {120}},\ \bibinfo
  {pages} {245102} (\bibinfo {year} {2016})}\BibitemShut {NoStop}%
\bibitem [{\citenamefont {Zhang}\ \emph {et~al.}(2009)\citenamefont {Zhang},
  \citenamefont {Park}, \citenamefont {Li}, \citenamefont {Lu}, \citenamefont
  {Zhang},\ and\ \citenamefont {Zhang}}]{Zhang_PhysRevLett_2009}%
  \BibitemOpen
  \bibfield  {author} {\bibinfo {author} {\bibfnamefont {S.}~\bibnamefont
  {Zhang}}, \bibinfo {author} {\bibfnamefont {Y.-S.}\ \bibnamefont {Park}},
  \bibinfo {author} {\bibfnamefont {J.}~\bibnamefont {Li}}, \bibinfo {author}
  {\bibfnamefont {X.}~\bibnamefont {Lu}}, \bibinfo {author} {\bibfnamefont
  {W.}~\bibnamefont {Zhang}}, \ and\ \bibinfo {author} {\bibfnamefont
  {X.}~\bibnamefont {Zhang}},\ }\bibfield  {title} {\enquote {\bibinfo {title}
  {Negative refractive index in chiral metamaterials},}\ }\href {\doibase
  10.1103/PhysRevLett.102.023901} {\bibfield  {journal} {\bibinfo  {journal}
  {Phys. Rev. Lett.}\ }\textbf {\bibinfo {volume} {102}},\ \bibinfo {pages}
  {023901} (\bibinfo {year} {2009})}\BibitemShut {NoStop}%
\bibitem [{\citenamefont {Fang}\ \emph {et~al.}(2017)\citenamefont {Fang},
  \citenamefont {Luan}, \citenamefont {Ma}, \citenamefont {Lv}, \citenamefont
  {Li}, \citenamefont {Zhu}, \citenamefont {Guan}, \citenamefont {Shi},\ and\
  \citenamefont {Cui}}]{Fang_JApplPhys_2017}%
  \BibitemOpen
  \bibfield  {author} {\bibinfo {author} {\bibfnamefont {S.}~\bibnamefont
  {Fang}}, \bibinfo {author} {\bibfnamefont {K.}~\bibnamefont {Luan}}, \bibinfo
  {author} {\bibfnamefont {H.~F.}\ \bibnamefont {Ma}}, \bibinfo {author}
  {\bibfnamefont {W.}~\bibnamefont {Lv}}, \bibinfo {author} {\bibfnamefont
  {Y.}~\bibnamefont {Li}}, \bibinfo {author} {\bibfnamefont {Z.}~\bibnamefont
  {Zhu}}, \bibinfo {author} {\bibfnamefont {C.}~\bibnamefont {Guan}}, \bibinfo
  {author} {\bibfnamefont {J.}~\bibnamefont {Shi}}, \ and\ \bibinfo {author}
  {\bibfnamefont {T.~J.}\ \bibnamefont {Cui}},\ }\bibfield  {title} {\enquote
  {\bibinfo {title} {Asymmetric transmission of linearly polarized waves in
  terahertz chiral metamaterials},}\ }\href {\doibase 10.1063/1.4974477}
  {\bibfield  {journal} {\bibinfo  {journal} {J. Appl. Phys.}\ }\textbf
  {\bibinfo {volume} {121}},\ \bibinfo {pages} {033103} (\bibinfo {year}
  {2017})}\BibitemShut {NoStop}%
\bibitem [{\citenamefont {Li}, \citenamefont {Mutlu},\ and\ \citenamefont
  {Ozbay}(2013)}]{Ozbay_JOpt_2013}%
  \BibitemOpen
  \bibfield  {author} {\bibinfo {author} {\bibfnamefont {Z.}~\bibnamefont
  {Li}}, \bibinfo {author} {\bibfnamefont {M.}~\bibnamefont {Mutlu}}, \ and\
  \bibinfo {author} {\bibfnamefont {E.}~\bibnamefont {Ozbay}},\ }\bibfield
  {title} {\enquote {\bibinfo {title} {Chiral metamaterials: from optical
  activity and negative refractive index to asymmetric transmission},}\ }\href
  {\doibase 10.1088/2040-8978/15/2/023001} {\bibfield  {journal} {\bibinfo
  {journal} {J. Opt.}\ }\textbf {\bibinfo {volume} {15}},\ \bibinfo {pages}
  {023001} (\bibinfo {year} {2013})}\BibitemShut {NoStop}%
\bibitem [{\citenamefont {Valev}\ \emph {et~al.}(2013)\citenamefont {Valev},
  \citenamefont {Baumberg}, \citenamefont {Sibilia},\ and\ \citenamefont
  {Verbiest}}]{Valev_AdvMater_2013}%
  \BibitemOpen
  \bibfield  {author} {\bibinfo {author} {\bibfnamefont {V.~K.}\ \bibnamefont
  {Valev}}, \bibinfo {author} {\bibfnamefont {J.~J.}\ \bibnamefont {Baumberg}},
  \bibinfo {author} {\bibfnamefont {C.}~\bibnamefont {Sibilia}}, \ and\
  \bibinfo {author} {\bibfnamefont {T.}~\bibnamefont {Verbiest}},\ }\bibfield
  {title} {\enquote {\bibinfo {title} {Chirality and chiroptical effects in
  plasmonic nanostructures: {Fundamentals}, recent progress, and outlook},}\
  }\href {\doibase 10.1002/adma.201205178} {\bibfield  {journal} {\bibinfo
  {journal} {Adv. Mater.}\ }\textbf {\bibinfo {volume} {25}},\ \bibinfo {pages}
  {2517--2534} (\bibinfo {year} {2013})}\BibitemShut {NoStop}%
\bibitem [{\citenamefont {Hentschel}\ \emph {et~al.}(2017)\citenamefont
  {Hentschel}, \citenamefont {Sch{\"a}ferling}, \citenamefont {Duan},
  \citenamefont {Giessen},\ and\ \citenamefont {Liu}}]{Hentschel_AAAS_2017}%
  \BibitemOpen
  \bibfield  {author} {\bibinfo {author} {\bibfnamefont {M.}~\bibnamefont
  {Hentschel}}, \bibinfo {author} {\bibfnamefont {M.}~\bibnamefont
  {Sch{\"a}ferling}}, \bibinfo {author} {\bibfnamefont {X.}~\bibnamefont
  {Duan}}, \bibinfo {author} {\bibfnamefont {H.}~\bibnamefont {Giessen}}, \
  and\ \bibinfo {author} {\bibfnamefont {N.}~\bibnamefont {Liu}},\ }\bibfield
  {title} {\enquote {\bibinfo {title} {Chiral plasmonics},}\ }\href {\doibase
  10.1126/sciadv.1602735} {\bibfield  {journal} {\bibinfo  {journal} {Sci.
  Adv.}\ }\textbf {\bibinfo {volume} {3}},\ \bibinfo {pages} {e1602735}
  (\bibinfo {year} {2017})}\BibitemShut {NoStop}%
\bibitem [{\citenamefont {Collins}\ \emph {et~al.}(2017)\citenamefont
  {Collins}, \citenamefont {Kuppe}, \citenamefont {Hooper}, \citenamefont
  {Sibilia}, \citenamefont {Centini},\ and\ \citenamefont
  {Valev}}]{Collins_AdvOptMat_2017}%
  \BibitemOpen
  \bibfield  {author} {\bibinfo {author} {\bibfnamefont {J.~T.}\ \bibnamefont
  {Collins}}, \bibinfo {author} {\bibfnamefont {C.}~\bibnamefont {Kuppe}},
  \bibinfo {author} {\bibfnamefont {D.~C.}\ \bibnamefont {Hooper}}, \bibinfo
  {author} {\bibfnamefont {C.}~\bibnamefont {Sibilia}}, \bibinfo {author}
  {\bibfnamefont {M.}~\bibnamefont {Centini}}, \ and\ \bibinfo {author}
  {\bibfnamefont {V.~K.}\ \bibnamefont {Valev}},\ }\bibfield  {title} {\enquote
  {\bibinfo {title} {Chirality and chiroptical effects in metal nanostructures:
  Fundamentals and current trends},}\ }\href {\doibase 10.1002/adom.201700182}
  {\bibfield  {journal} {\bibinfo  {journal} {Adv. Opt. Mater.}\ }\textbf
  {\bibinfo {volume} {5}},\ \bibinfo {pages} {1700182} (\bibinfo {year}
  {2017})}\BibitemShut {NoStop}%
\bibitem [{\citenamefont {Arteaga}\ \emph {et~al.}(2016)\citenamefont
  {Arteaga}, \citenamefont {{Sancho-Parramon}}, \citenamefont {Nichols},
  \citenamefont {Maoz}, \citenamefont {Canillas}, \citenamefont {Bosch},
  \citenamefont {Markovich},\ and\ \citenamefont
  {Kahr}}]{Arteaga_OptExpress_2016}%
  \BibitemOpen
  \bibfield  {author} {\bibinfo {author} {\bibfnamefont {O.}~\bibnamefont
  {Arteaga}}, \bibinfo {author} {\bibfnamefont {J.}~\bibnamefont
  {{Sancho-Parramon}}}, \bibinfo {author} {\bibfnamefont {S.}~\bibnamefont
  {Nichols}}, \bibinfo {author} {\bibfnamefont {B.~M.}\ \bibnamefont {Maoz}},
  \bibinfo {author} {\bibfnamefont {A.}~\bibnamefont {Canillas}}, \bibinfo
  {author} {\bibfnamefont {S.}~\bibnamefont {Bosch}}, \bibinfo {author}
  {\bibfnamefont {G.}~\bibnamefont {Markovich}}, \ and\ \bibinfo {author}
  {\bibfnamefont {B.}~\bibnamefont {Kahr}},\ }\bibfield  {title} {\enquote
  {\bibinfo {title} {Relation between {2D/3D} chirality and the appearance of
  chiroptical effects in real nanostructures},}\ }\href {\doibase
  10.1364/OE.24.002242} {\bibfield  {journal} {\bibinfo  {journal} {Opt.
  Express}\ }\textbf {\bibinfo {volume} {24}},\ \bibinfo {pages} {2242--2252}
  (\bibinfo {year} {2016})}\BibitemShut {NoStop}%
\bibitem [{\citenamefont {Gansel}\ \emph {et~al.}(2009)\citenamefont {Gansel},
  \citenamefont {Thiel}, \citenamefont {Rill}, \citenamefont {Decker},
  \citenamefont {Bade}, \citenamefont {Saile}, \citenamefont {{von Freymann}},
  \citenamefont {Linden},\ and\ \citenamefont {Wegener}}]{Gansel_Science_2009}%
  \BibitemOpen
  \bibfield  {author} {\bibinfo {author} {\bibfnamefont {J.~K.}\ \bibnamefont
  {Gansel}}, \bibinfo {author} {\bibfnamefont {M.}~\bibnamefont {Thiel}},
  \bibinfo {author} {\bibfnamefont {M.~S.}\ \bibnamefont {Rill}}, \bibinfo
  {author} {\bibfnamefont {M.}~\bibnamefont {Decker}}, \bibinfo {author}
  {\bibfnamefont {K.}~\bibnamefont {Bade}}, \bibinfo {author} {\bibfnamefont
  {V.}~\bibnamefont {Saile}}, \bibinfo {author} {\bibfnamefont
  {G.}~\bibnamefont {{von Freymann}}}, \bibinfo {author} {\bibfnamefont
  {S.}~\bibnamefont {Linden}}, \ and\ \bibinfo {author} {\bibfnamefont
  {M.}~\bibnamefont {Wegener}},\ }\bibfield  {title} {\enquote {\bibinfo
  {title} {Gold helix photonic metamaterial as broadband circular polarizer},}\
  }\href {\doibase 10.1126/science.1177031} {\bibfield  {journal} {\bibinfo
  {journal} {Science}\ }\textbf {\bibinfo {volume} {325}},\ \bibinfo {pages}
  {1513--1515} (\bibinfo {year} {2009})}\BibitemShut {NoStop}%
\bibitem [{\citenamefont {Thiel}\ \emph {et~al.}(2009)\citenamefont {Thiel},
  \citenamefont {Rill}, \citenamefont {{von Freymann}},\ and\ \citenamefont
  {Wegener}}]{Wegener_AdvMater_2009}%
  \BibitemOpen
  \bibfield  {author} {\bibinfo {author} {\bibfnamefont {M.}~\bibnamefont
  {Thiel}}, \bibinfo {author} {\bibfnamefont {M.~S.}\ \bibnamefont {Rill}},
  \bibinfo {author} {\bibfnamefont {G.}~\bibnamefont {{von Freymann}}}, \ and\
  \bibinfo {author} {\bibfnamefont {M.}~\bibnamefont {Wegener}},\ }\bibfield
  {title} {\enquote {\bibinfo {title} {Three-dimensional bi-chiral photonic
  crystals},}\ }\href {\doibase 10.1002/adma.200901601} {\bibfield  {journal}
  {\bibinfo  {journal} {Adv. Mater.}\ }\textbf {\bibinfo {volume} {21}},\
  \bibinfo {pages} {4680--4682} (\bibinfo {year} {2009})}\BibitemShut {NoStop}%
\bibitem [{\citenamefont {Gansel}\ \emph {et~al.}(2012)\citenamefont {Gansel},
  \citenamefont {Latzel}, \citenamefont {Fr\"olich}, \citenamefont {Kaschke},
  \citenamefont {Thiel},\ and\ \citenamefont
  {Wegener}}]{Gansel_ApplPhysLett_2012}%
  \BibitemOpen
  \bibfield  {author} {\bibinfo {author} {\bibfnamefont {J.~K.}\ \bibnamefont
  {Gansel}}, \bibinfo {author} {\bibfnamefont {M.}~\bibnamefont {Latzel}},
  \bibinfo {author} {\bibfnamefont {A.}~\bibnamefont {Fr\"olich}}, \bibinfo
  {author} {\bibfnamefont {J.}~\bibnamefont {Kaschke}}, \bibinfo {author}
  {\bibfnamefont {M.}~\bibnamefont {Thiel}}, \ and\ \bibinfo {author}
  {\bibfnamefont {M.}~\bibnamefont {Wegener}},\ }\bibfield  {title} {\enquote
  {\bibinfo {title} {Tapered gold-helix metamaterials as improved circular
  polarizers},}\ }\href {\doibase 10.1063/1.3693181} {\bibfield  {journal}
  {\bibinfo  {journal} {Appl. Phys. Lett.}\ }\textbf {\bibinfo {volume}
  {100}},\ \bibinfo {pages} {101109} (\bibinfo {year} {2012})}\BibitemShut
  {NoStop}%
\bibitem [{\citenamefont {Sakellari}\ \emph {et~al.}(2017)\citenamefont
  {Sakellari}, \citenamefont {Yin}, \citenamefont {Nesterov}, \citenamefont
  {Terzaki}, \citenamefont {Xomalis},\ and\ \citenamefont
  {Farsari}}]{Sakellari_AdvOptMater_2017}%
  \BibitemOpen
  \bibfield  {author} {\bibinfo {author} {\bibfnamefont {I.}~\bibnamefont
  {Sakellari}}, \bibinfo {author} {\bibfnamefont {X.}~\bibnamefont {Yin}},
  \bibinfo {author} {\bibfnamefont {M.~L.}\ \bibnamefont {Nesterov}}, \bibinfo
  {author} {\bibfnamefont {K.}~\bibnamefont {Terzaki}}, \bibinfo {author}
  {\bibfnamefont {A.}~\bibnamefont {Xomalis}}, \ and\ \bibinfo {author}
  {\bibfnamefont {M.}~\bibnamefont {Farsari}},\ }\bibfield  {title} {\enquote
  {\bibinfo {title} {{3D} chiral plasmonic metamaterials fabricated by direct
  laser writing: The twisted omega particle},}\ }\href {\doibase
  10.1002/adom.201700200} {\bibfield  {journal} {\bibinfo  {journal} {Adv. Opt.
  Mater.}\ }\textbf {\bibinfo {volume} {5}},\ \bibinfo {pages} {1700200}
  (\bibinfo {year} {2017})}\BibitemShut {NoStop}%
\bibitem [{\citenamefont {Tsutsumi}\ \emph {et~al.}(2017)\citenamefont
  {Tsutsumi}, \citenamefont {Fukuda}, \citenamefont {Nakamura}, \citenamefont
  {Kinashi},\ and\ \citenamefont {Sakai}}]{Tsutsumi2017}%
  \BibitemOpen
  \bibfield  {author} {\bibinfo {author} {\bibfnamefont {N.}~\bibnamefont
  {Tsutsumi}}, \bibinfo {author} {\bibfnamefont {A.}~\bibnamefont {Fukuda}},
  \bibinfo {author} {\bibfnamefont {R.}~\bibnamefont {Nakamura}}, \bibinfo
  {author} {\bibfnamefont {K.}~\bibnamefont {Kinashi}}, \ and\ \bibinfo
  {author} {\bibfnamefont {W.}~\bibnamefont {Sakai}},\ }\bibfield  {title}
  {\enquote {\bibinfo {title} {Fabrication of three-dimensional microstructures
  in positive photoresist through two-photon direct laser writing},}\ }\href
  {\doibase 10.1007/s00339-017-1170-4} {\bibfield  {journal} {\bibinfo
  {journal} {Appl. Phys. A}\ }\textbf {\bibinfo {volume} {123}},\ \bibinfo
  {pages} {553} (\bibinfo {year} {2017})}\BibitemShut {NoStop}%
\bibitem [{\citenamefont {Golod}\ \emph {et~al.}(2018)\citenamefont {Golod},
  \citenamefont {Seyfi}, \citenamefont {Buldygin}, \citenamefont {Gayduk},\
  and\ \citenamefont {Prinz}}]{Golod_AdvOptMater_2018}%
  \BibitemOpen
  \bibfield  {author} {\bibinfo {author} {\bibfnamefont {S.~V.}\ \bibnamefont
  {Golod}}, \bibinfo {author} {\bibfnamefont {V.~A.}\ \bibnamefont {Seyfi}},
  \bibinfo {author} {\bibfnamefont {A.~F.}\ \bibnamefont {Buldygin}}, \bibinfo
  {author} {\bibfnamefont {A.~E.}\ \bibnamefont {Gayduk}}, \ and\ \bibinfo
  {author} {\bibfnamefont {V.~Y.}\ \bibnamefont {Prinz}},\ }\bibfield  {title}
  {\enquote {\bibinfo {title} {Large-area {3D}-printed chiral metasurface
  composed of metal helices},}\ }\href {\doibase 10.1002/adom.201800424}
  {\bibfield  {journal} {\bibinfo  {journal} {Adv. Opt. Mater.}\ }\textbf
  {\bibinfo {volume} {6}},\ \bibinfo {pages} {1800424} (\bibinfo {year}
  {2018})}\BibitemShut {NoStop}%
\bibitem [{\citenamefont {Zhu}\ \emph {et~al.}(2018)\citenamefont {Zhu},
  \citenamefont {Chen}, \citenamefont {Zaidi}, \citenamefont {Huang},
  \citenamefont {Khorasaninejad}, \citenamefont {Sanjeev}, \citenamefont
  {Qiu},\ and\ \citenamefont {Capasso}}]{Zhu_LightSciAppl_2018}%
  \BibitemOpen
  \bibfield  {author} {\bibinfo {author} {\bibfnamefont {A.~Y.}\ \bibnamefont
  {Zhu}}, \bibinfo {author} {\bibfnamefont {W.~T.}\ \bibnamefont {Chen}},
  \bibinfo {author} {\bibfnamefont {A.}~\bibnamefont {Zaidi}}, \bibinfo
  {author} {\bibfnamefont {Y.-W.}\ \bibnamefont {Huang}}, \bibinfo {author}
  {\bibfnamefont {M.}~\bibnamefont {Khorasaninejad}}, \bibinfo {author}
  {\bibfnamefont {V.}~\bibnamefont {Sanjeev}}, \bibinfo {author} {\bibfnamefont
  {C.-W.}\ \bibnamefont {Qiu}}, \ and\ \bibinfo {author} {\bibfnamefont
  {F.}~\bibnamefont {Capasso}},\ }\bibfield  {title} {\enquote {\bibinfo
  {title} {Giant intrinsic chiro-optical activity in planar dielectric
  nanostructures},}\ }\href {\doibase 10.1038/lsa.2017.158} {\bibfield
  {journal} {\bibinfo  {journal} {Light Sci. Appl.}\ }\textbf {\bibinfo
  {volume} {7}},\ \bibinfo {pages} {17158} (\bibinfo {year}
  {2018})}\BibitemShut {NoStop}%
\bibitem [{\citenamefont {Gorkunov}\ \emph {et~al.}(2018)\citenamefont
  {Gorkunov}, \citenamefont {Rogov}, \citenamefont {Kondratov}, \citenamefont
  {Artemov}, \citenamefont {Gainutdinov},\ and\ \citenamefont
  {Ezhov}}]{Gorkunov_SciRep_2018}%
  \BibitemOpen
  \bibfield  {author} {\bibinfo {author} {\bibfnamefont {M.~V.}\ \bibnamefont
  {Gorkunov}}, \bibinfo {author} {\bibfnamefont {O.~Y.}\ \bibnamefont {Rogov}},
  \bibinfo {author} {\bibfnamefont {A.~V.}\ \bibnamefont {Kondratov}}, \bibinfo
  {author} {\bibfnamefont {V.~V.}\ \bibnamefont {Artemov}}, \bibinfo {author}
  {\bibfnamefont {R.~V.}\ \bibnamefont {Gainutdinov}}, \ and\ \bibinfo {author}
  {\bibfnamefont {A.~A.}\ \bibnamefont {Ezhov}},\ }\bibfield  {title} {\enquote
  {\bibinfo {title} {Chiral visible light metasurface patterned in
  monocrystalline silicon by focused ion beam},}\ }\href {\doibase
  10.1038/s41598-018-29977-4} {\bibfield  {journal} {\bibinfo  {journal} {Sci.
  Rep.}\ }\textbf {\bibinfo {volume} {8}},\ \bibinfo {pages} {11623} (\bibinfo
  {year} {2018})}\BibitemShut {NoStop}%
\bibitem [{\citenamefont {Wu}\ \emph {et~al.}(2014)\citenamefont {Wu},
  \citenamefont {Ng}, \citenamefont {Liang}, \citenamefont {Breese},
  \citenamefont {Hong}, \citenamefont {Maier}, \citenamefont {Moser},\ and\
  \citenamefont {Hess}}]{Wu_PhysRevApplied_2014}%
  \BibitemOpen
  \bibfield  {author} {\bibinfo {author} {\bibfnamefont {J.}~\bibnamefont
  {Wu}}, \bibinfo {author} {\bibfnamefont {B.}~\bibnamefont {Ng}}, \bibinfo
  {author} {\bibfnamefont {H.}~\bibnamefont {Liang}}, \bibinfo {author}
  {\bibfnamefont {M.~B.~H.}\ \bibnamefont {Breese}}, \bibinfo {author}
  {\bibfnamefont {M.}~\bibnamefont {Hong}}, \bibinfo {author} {\bibfnamefont
  {S.~A.}\ \bibnamefont {Maier}}, \bibinfo {author} {\bibfnamefont {H.~O.}\
  \bibnamefont {Moser}}, \ and\ \bibinfo {author} {\bibfnamefont
  {O.}~\bibnamefont {Hess}},\ }\bibfield  {title} {\enquote {\bibinfo {title}
  {Chiral metafoils for terahertz broadband high-contrast flexible circular
  polarizers},}\ }\href {\doibase 10.1103/PhysRevApplied.2.014005} {\bibfield
  {journal} {\bibinfo  {journal} {Phys. Rev. Applied}\ }\textbf {\bibinfo
  {volume} {2}},\ \bibinfo {pages} {014005} (\bibinfo {year}
  {2014})}\BibitemShut {NoStop}%
\bibitem [{\citenamefont {Park}\ \emph {et~al.}(2014)\citenamefont {Park},
  \citenamefont {Kim}, \citenamefont {Kim}, \citenamefont {Kim},\ and\
  \citenamefont {Min}}]{Park_NaCommun_2014}%
  \BibitemOpen
  \bibfield  {author} {\bibinfo {author} {\bibfnamefont {H.~S.}\ \bibnamefont
  {Park}}, \bibinfo {author} {\bibfnamefont {T.-T.}\ \bibnamefont {Kim}},
  \bibinfo {author} {\bibfnamefont {H.-D.}\ \bibnamefont {Kim}}, \bibinfo
  {author} {\bibfnamefont {K.}~\bibnamefont {Kim}}, \ and\ \bibinfo {author}
  {\bibfnamefont {B.}~\bibnamefont {Min}},\ }\bibfield  {title} {\enquote
  {\bibinfo {title} {Nondispersive optical activity of meshed helical
  metamaterials},}\ }\href {\doibase 10.1038/ncomms6435} {\bibfield  {journal}
  {\bibinfo  {journal} {Nat. Commun.}\ }\textbf {\bibinfo {volume} {5}},\
  \bibinfo {pages} {5435} (\bibinfo {year} {2014})}\BibitemShut {NoStop}%
\bibitem [{\citenamefont {Fasold}\ \emph {et~al.}(2018)\citenamefont {Fasold},
  \citenamefont {Lin{\ss}}, \citenamefont {Kawde}, \citenamefont {Falkner},
  \citenamefont {Decker}, \citenamefont {Pertsch},\ and\ \citenamefont
  {Staude}}]{Fasold_acsphotonics_2018}%
  \BibitemOpen
  \bibfield  {author} {\bibinfo {author} {\bibfnamefont {S.}~\bibnamefont
  {Fasold}}, \bibinfo {author} {\bibfnamefont {S.}~\bibnamefont {Lin{\ss}}},
  \bibinfo {author} {\bibfnamefont {T.}~\bibnamefont {Kawde}}, \bibinfo
  {author} {\bibfnamefont {M.}~\bibnamefont {Falkner}}, \bibinfo {author}
  {\bibfnamefont {M.}~\bibnamefont {Decker}}, \bibinfo {author} {\bibfnamefont
  {T.}~\bibnamefont {Pertsch}}, \ and\ \bibinfo {author} {\bibfnamefont
  {I.}~\bibnamefont {Staude}},\ }\bibfield  {title} {\enquote {\bibinfo {title}
  {Disorder-enabled pure chirality in bilayer plasmonic metasurfaces},}\ }\href
  {\doibase 10.1021/acsphotonics.7b01460} {\bibfield  {journal} {\bibinfo
  {journal} {ACS Photonics}\ }\textbf {\bibinfo {volume} {5}},\ \bibinfo
  {pages} {1773--1778} (\bibinfo {year} {2018})}\BibitemShut {NoStop}%
\bibitem [{\citenamefont {Hentschel}\ \emph {et~al.}(2012)\citenamefont
  {Hentschel}, \citenamefont {Wu}, \citenamefont {Sch\"aferling}, \citenamefont
  {Bai}, \citenamefont {Li},\ and\ \citenamefont
  {Giessen}}]{Hentschel_ACSNano_2012}%
  \BibitemOpen
  \bibfield  {author} {\bibinfo {author} {\bibfnamefont {M.}~\bibnamefont
  {Hentschel}}, \bibinfo {author} {\bibfnamefont {L.}~\bibnamefont {Wu}},
  \bibinfo {author} {\bibfnamefont {M.}~\bibnamefont {Sch\"aferling}}, \bibinfo
  {author} {\bibfnamefont {P.}~\bibnamefont {Bai}}, \bibinfo {author}
  {\bibfnamefont {E.~P.}\ \bibnamefont {Li}}, \ and\ \bibinfo {author}
  {\bibfnamefont {H.}~\bibnamefont {Giessen}},\ }\bibfield  {title} {\enquote
  {\bibinfo {title} {Optical properties of chiral three-dimensional plasmonic
  oligomers at the onset of charge-transfer plasmons},}\ }\href {\doibase
  10.1021/nn304283y} {\bibfield  {journal} {\bibinfo  {journal} {ACS Nano}\
  }\textbf {\bibinfo {volume} {6}},\ \bibinfo {pages} {10355--10365} (\bibinfo
  {year} {2012})}\BibitemShut {NoStop}%
\bibitem [{\citenamefont {Slaughter}\ \emph {et~al.}(2010)\citenamefont
  {Slaughter}, \citenamefont {Wu}, \citenamefont {Willingham}, \citenamefont
  {Nordlander},\ and\ \citenamefont {Link}}]{Slaughter_ACSNano_2010}%
  \BibitemOpen
  \bibfield  {author} {\bibinfo {author} {\bibfnamefont {L.~S.}\ \bibnamefont
  {Slaughter}}, \bibinfo {author} {\bibfnamefont {Y.}~\bibnamefont {Wu}},
  \bibinfo {author} {\bibfnamefont {B.~A.}\ \bibnamefont {Willingham}},
  \bibinfo {author} {\bibfnamefont {P.}~\bibnamefont {Nordlander}}, \ and\
  \bibinfo {author} {\bibfnamefont {S.}~\bibnamefont {Link}},\ }\bibfield
  {title} {\enquote {\bibinfo {title} {Effects of symmetry breaking and
  conductive contact on the plasmon coupling in gold nanorod dimers},}\ }\href
  {\doibase 10.1021/nn1011144} {\bibfield  {journal} {\bibinfo  {journal} {ACS
  Nano}\ }\textbf {\bibinfo {volume} {4}},\ \bibinfo {pages} {4657--4666}
  (\bibinfo {year} {2010})}\BibitemShut {NoStop}%
\bibitem [{\citenamefont {Svirko}, \citenamefont {Zheludev},\ and\
  \citenamefont {Osipov}(2001)}]{Svirko_ApplPhysLett_2001}%
  \BibitemOpen
  \bibfield  {author} {\bibinfo {author} {\bibfnamefont {Y.}~\bibnamefont
  {Svirko}}, \bibinfo {author} {\bibfnamefont {N.}~\bibnamefont {Zheludev}}, \
  and\ \bibinfo {author} {\bibfnamefont {M.}~\bibnamefont {Osipov}},\
  }\bibfield  {title} {\enquote {\bibinfo {title} {Layered chiral metallic
  microstructures with inductive coupling},}\ }\href {\doibase
  10.1063/1.1342210} {\bibfield  {journal} {\bibinfo  {journal} {Appl. Phys.
  Lett.}\ }\textbf {\bibinfo {volume} {78}},\ \bibinfo {pages} {498--500}
  (\bibinfo {year} {2001})}\BibitemShut {NoStop}%
\bibitem [{\citenamefont {Menzel}, \citenamefont {Rockstuhl},\ and\
  \citenamefont {Lederer}(2010)}]{Menzel_PhysRevA_2010}%
  \BibitemOpen
  \bibfield  {author} {\bibinfo {author} {\bibfnamefont {C.}~\bibnamefont
  {Menzel}}, \bibinfo {author} {\bibfnamefont {C.}~\bibnamefont {Rockstuhl}}, \
  and\ \bibinfo {author} {\bibfnamefont {F.}~\bibnamefont {Lederer}},\
  }\bibfield  {title} {\enquote {\bibinfo {title} {Advanced {Jones} calculus
  for the classification of periodic metamaterials},}\ }\href {\doibase
  10.1103/PhysRevA.82.053811} {\bibfield  {journal} {\bibinfo  {journal} {Phys.
  Rev. A}\ }\textbf {\bibinfo {volume} {82}},\ \bibinfo {pages} {053811}
  (\bibinfo {year} {2010})}\BibitemShut {NoStop}%
\bibitem [{\citenamefont {Pavlov}\ \emph {et~al.}(2013)\citenamefont {Pavlov},
  \citenamefont {Klimov}, \citenamefont {Vladimorova},\ and\ \citenamefont
  {Zadkov}}]{Pavlov_2013}%
  \BibitemOpen
  \bibfield  {author} {\bibinfo {author} {\bibfnamefont {A.~A.}\ \bibnamefont
  {Pavlov}}, \bibinfo {author} {\bibfnamefont {V.~V.}\ \bibnamefont {Klimov}},
  \bibinfo {author} {\bibfnamefont {Y.~V.}\ \bibnamefont {Vladimorova}}, \ and\
  \bibinfo {author} {\bibfnamefont {V.~N.}\ \bibnamefont {Zadkov}},\ }\bibfield
   {title} {\enquote {\bibinfo {title} {Analysis of optical properties of
  planar metamaterials by calculating multipole moments of their constituent
  meta-atoms},}\ }\href {\doibase 10.1070/qe2013v043n05abeh015057} {\bibfield
  {journal} {\bibinfo  {journal} {Quantum Electron.}\ }\textbf {\bibinfo
  {volume} {43}},\ \bibinfo {pages} {496--501} (\bibinfo {year}
  {2013})}\BibitemShut {NoStop}%
\bibitem [{\citenamefont {Qu}\ \emph {et~al.}(2018)\citenamefont {Qu},
  \citenamefont {Zhang}, \citenamefont {Wang}, \citenamefont {Li},
  \citenamefont {Ullah}, \citenamefont {Aba}, \citenamefont {Wang},
  \citenamefont {Fu},\ and\ \citenamefont {Zhang}}]{Qu_AnnPhys_2018}%
  \BibitemOpen
  \bibfield  {author} {\bibinfo {author} {\bibfnamefont {Y.}~\bibnamefont
  {Qu}}, \bibinfo {author} {\bibfnamefont {Y.}~\bibnamefont {Zhang}}, \bibinfo
  {author} {\bibfnamefont {F.}~\bibnamefont {Wang}}, \bibinfo {author}
  {\bibfnamefont {H.}~\bibnamefont {Li}}, \bibinfo {author} {\bibfnamefont
  {H.}~\bibnamefont {Ullah}}, \bibinfo {author} {\bibfnamefont
  {T.}~\bibnamefont {Aba}}, \bibinfo {author} {\bibfnamefont {Y.}~\bibnamefont
  {Wang}}, \bibinfo {author} {\bibfnamefont {T.}~\bibnamefont {Fu}}, \ and\
  \bibinfo {author} {\bibfnamefont {Z.}~\bibnamefont {Zhang}},\ }\bibfield
  {title} {\enquote {\bibinfo {title} {A general mechanism for achieving
  circular dichroism in a chiral plasmonic system},}\ }\href {\doibase
  10.1002/andp.201800142} {\bibfield  {journal} {\bibinfo  {journal} {Ann.
  Phys. (Berl.)}\ }\textbf {\bibinfo {volume} {530}},\ \bibinfo {pages}
  {1800142} (\bibinfo {year} {2018})}\BibitemShut {NoStop}%
\bibitem [{\citenamefont {Zhang}, \citenamefont {Lu},\ and\ \citenamefont
  {Zheng}(2018)}]{Zhang_JOptSocAmB_2018}%
  \BibitemOpen
  \bibfield  {author} {\bibinfo {author} {\bibfnamefont {M.}~\bibnamefont
  {Zhang}}, \bibinfo {author} {\bibfnamefont {Q.}~\bibnamefont {Lu}}, \ and\
  \bibinfo {author} {\bibfnamefont {H.}~\bibnamefont {Zheng}},\ }\bibfield
  {title} {\enquote {\bibinfo {title} {Tunable circular dichroism created by
  surface plasmons in bilayer twisted tetramer nanostructure arrays},}\ }\href
  {\doibase 10.1364/JOSAB.35.000689} {\bibfield  {journal} {\bibinfo  {journal}
  {J. Opt. Soc. Am. B}\ }\textbf {\bibinfo {volume} {35}},\ \bibinfo {pages}
  {689--693} (\bibinfo {year} {2018})}\BibitemShut {NoStop}%
\bibitem [{\citenamefont {Sperrhake}\ \emph {et~al.}(2019)\citenamefont
  {Sperrhake}, \citenamefont {Decker}, \citenamefont {Falkner}, \citenamefont
  {Fasold}, \citenamefont {Kaiser}, \citenamefont {Staude},\ and\ \citenamefont
  {Pertsch}}]{Sperrhake_OptExpress_2019}%
  \BibitemOpen
  \bibfield  {author} {\bibinfo {author} {\bibfnamefont {J.}~\bibnamefont
  {Sperrhake}}, \bibinfo {author} {\bibfnamefont {M.}~\bibnamefont {Decker}},
  \bibinfo {author} {\bibfnamefont {M.}~\bibnamefont {Falkner}}, \bibinfo
  {author} {\bibfnamefont {S.}~\bibnamefont {Fasold}}, \bibinfo {author}
  {\bibfnamefont {T.}~\bibnamefont {Kaiser}}, \bibinfo {author} {\bibfnamefont
  {I.}~\bibnamefont {Staude}}, \ and\ \bibinfo {author} {\bibfnamefont
  {T.}~\bibnamefont {Pertsch}},\ }\bibfield  {title} {\enquote {\bibinfo
  {title} {Analyzing the polarization response of a chiral metasurface stack by
  semi-analytic modeling},}\ }\href {\doibase 10.1364/OE.27.001236} {\bibfield
  {journal} {\bibinfo  {journal} {Opt. Express}\ }\textbf {\bibinfo {volume}
  {27}},\ \bibinfo {pages} {1236--1248} (\bibinfo {year} {2019})}\BibitemShut
  {NoStop}%
\bibitem [{\citenamefont {Headland}\ \emph {et~al.}(2016)\citenamefont
  {Headland}, \citenamefont {Withayachumnankul}, \citenamefont {Webb},
  \citenamefont {{Ebendorff-Heidepriem}}, \citenamefont {Luiten},\ and\
  \citenamefont {Abbott}}]{Headland_OptExpress_2016}%
  \BibitemOpen
  \bibfield  {author} {\bibinfo {author} {\bibfnamefont {D.}~\bibnamefont
  {Headland}}, \bibinfo {author} {\bibfnamefont {W.}~\bibnamefont
  {Withayachumnankul}}, \bibinfo {author} {\bibfnamefont {M.}~\bibnamefont
  {Webb}}, \bibinfo {author} {\bibfnamefont {H.}~\bibnamefont
  {{Ebendorff-Heidepriem}}}, \bibinfo {author} {\bibfnamefont {A.}~\bibnamefont
  {Luiten}}, \ and\ \bibinfo {author} {\bibfnamefont {D.}~\bibnamefont
  {Abbott}},\ }\bibfield  {title} {\enquote {\bibinfo {title} {Analysis of
  {3D}-printed metal for rapid-prototyped reflective terahertz optics},}\
  }\href {\doibase 10.1364/OE.24.017384} {\bibfield  {journal} {\bibinfo
  {journal} {Opt. Express}\ }\textbf {\bibinfo {volume} {24}},\ \bibinfo
  {pages} {17384--17396} (\bibinfo {year} {2016})}\BibitemShut {NoStop}%
\bibitem [{\citenamefont {Camposeo}\ \emph {et~al.}(2019)\citenamefont
  {Camposeo}, \citenamefont {Persano}, \citenamefont {Farsari},\ and\
  \citenamefont {Pisignano}}]{Camposeo_AdvOptMater_2019}%
  \BibitemOpen
  \bibfield  {author} {\bibinfo {author} {\bibfnamefont {A.}~\bibnamefont
  {Camposeo}}, \bibinfo {author} {\bibfnamefont {L.}~\bibnamefont {Persano}},
  \bibinfo {author} {\bibfnamefont {M.}~\bibnamefont {Farsari}}, \ and\
  \bibinfo {author} {\bibfnamefont {D.}~\bibnamefont {Pisignano}},\ }\bibfield
  {title} {\enquote {\bibinfo {title} {Additive manufacturing: {Applications}
  and directions in photonics and optoelectronics},}\ }\href {\doibase
  10.1002/adom.201800419} {\bibfield  {journal} {\bibinfo  {journal} {Adv. Opt.
  Mater.}\ }\textbf {\bibinfo {volume} {7}},\ \bibinfo {pages} {1800419}
  (\bibinfo {year} {2019})}\BibitemShut {NoStop}%
\bibitem [{\citenamefont {Wu}\ \emph {et~al.}(2019)\citenamefont {Wu},
  \citenamefont {Xu}, \citenamefont {Zinenko}, \citenamefont {Yachin},
  \citenamefont {Prosvirnin},\ and\ \citenamefont {Tuz}}]{Wu_OL_2019}%
  \BibitemOpen
  \bibfield  {author} {\bibinfo {author} {\bibfnamefont {S.}~\bibnamefont
  {Wu}}, \bibinfo {author} {\bibfnamefont {S.}~\bibnamefont {Xu}}, \bibinfo
  {author} {\bibfnamefont {T.~L.}\ \bibnamefont {Zinenko}}, \bibinfo {author}
  {\bibfnamefont {V.~V.}\ \bibnamefont {Yachin}}, \bibinfo {author}
  {\bibfnamefont {S.~L.}\ \bibnamefont {Prosvirnin}}, \ and\ \bibinfo {author}
  {\bibfnamefont {V.~R.}\ \bibnamefont {Tuz}},\ }\bibfield  {title} {\enquote
  {\bibinfo {title} {{3D}-printed chiral metasurface as a dichroic dual-band
  polarization converter},}\ }\href {\doibase 10.1364/OL.44.001056} {\bibfield
  {journal} {\bibinfo  {journal} {Opt. Lett.}\ }\textbf {\bibinfo {volume}
  {44}},\ \bibinfo {pages} {1056--1059} (\bibinfo {year} {2019})}\BibitemShut
  {NoStop}%
\bibitem [{\citenamefont {Yachin}\ and\ \citenamefont
  {Yasumoto}(2007)}]{Yachin_JOSAA_2007}%
  \BibitemOpen
  \bibfield  {author} {\bibinfo {author} {\bibfnamefont {V.}~\bibnamefont
  {Yachin}}\ and\ \bibinfo {author} {\bibfnamefont {K.}~\bibnamefont
  {Yasumoto}},\ }\bibfield  {title} {\enquote {\bibinfo {title} {Method of
  integral functionals for electromagnetic wave scattering from a
  double-periodic magnetodielectric layer},}\ }\href {\doibase
  10.1364/JOSAA.24.003606} {\bibfield  {journal} {\bibinfo  {journal} {J. Opt.
  Soc. Am. A}\ }\textbf {\bibinfo {volume} {24}},\ \bibinfo {pages}
  {3606--3618} (\bibinfo {year} {2007})}\BibitemShut {NoStop}%
\bibitem [{\citenamefont {Li}(1996)}]{Li_JOSAA_1996}%
  \BibitemOpen
  \bibfield  {author} {\bibinfo {author} {\bibfnamefont {L.}~\bibnamefont
  {Li}},\ }\bibfield  {title} {\enquote {\bibinfo {title} {Formulation and
  comparison of two recursive matrix algorithms for modeling layered
  diffraction gratings},}\ }\href {\doibase 10.1364/JOSAA.13.001024} {\bibfield
   {journal} {\bibinfo  {journal} {J. Opt. Soc. Am. A}\ }\textbf {\bibinfo
  {volume} {13}},\ \bibinfo {pages} {1024--1035} (\bibinfo {year}
  {1996})}\BibitemShut {NoStop}%
\bibitem [{\citenamefont {Moharam}\ \emph {et~al.}(1995)\citenamefont
  {Moharam}, \citenamefont {Pommet}, \citenamefont {Grann},\ and\ \citenamefont
  {Gaylord}}]{Moharam_JOSAA_95}%
  \BibitemOpen
  \bibfield  {author} {\bibinfo {author} {\bibfnamefont {M.~G.}\ \bibnamefont
  {Moharam}}, \bibinfo {author} {\bibfnamefont {D.~A.}\ \bibnamefont {Pommet}},
  \bibinfo {author} {\bibfnamefont {E.~B.}\ \bibnamefont {Grann}}, \ and\
  \bibinfo {author} {\bibfnamefont {T.~K.}\ \bibnamefont {Gaylord}},\
  }\bibfield  {title} {\enquote {\bibinfo {title} {Stable implementation of the
  rigorous coupled-wave analysis for surface-relief gratings: {Enhanced}
  transmittance matrix approach},}\ }\href {\doibase 10.1364/JOSAA.12.001077}
  {\bibfield  {journal} {\bibinfo  {journal} {J. Opt. Soc. Am. A}\ }\textbf
  {\bibinfo {volume} {12}},\ \bibinfo {pages} {1077--1086} (\bibinfo {year}
  {1995})}\BibitemShut {NoStop}%
\bibitem [{\citenamefont {Li}(2003)}]{Li_JOSAA_2003}%
  \BibitemOpen
  \bibfield  {author} {\bibinfo {author} {\bibfnamefont {L.}~\bibnamefont
  {Li}},\ }\bibfield  {title} {\enquote {\bibinfo {title} {Note on the
  {$S$-matrix} propagation algorithm},}\ }\href {\doibase
  10.1364/JOSAA.20.000655} {\bibfield  {journal} {\bibinfo  {journal} {J. Opt.
  Soc. Am. A}\ }\textbf {\bibinfo {volume} {20}},\ \bibinfo {pages} {655--660}
  (\bibinfo {year} {2003})}\BibitemShut {NoStop}%
\bibitem [{\citenamefont {Menzel}\ \emph {et~al.}(2010)\citenamefont {Menzel},
  \citenamefont {Helgert}, \citenamefont {Rockstuhl}, \citenamefont {Kley},
  \citenamefont {T\"unnermann}, \citenamefont {Pertsch},\ and\ \citenamefont
  {Lederer}}]{Menzel_PhysRevLett_2010}%
  \BibitemOpen
  \bibfield  {author} {\bibinfo {author} {\bibfnamefont {C.}~\bibnamefont
  {Menzel}}, \bibinfo {author} {\bibfnamefont {C.}~\bibnamefont {Helgert}},
  \bibinfo {author} {\bibfnamefont {C.}~\bibnamefont {Rockstuhl}}, \bibinfo
  {author} {\bibfnamefont {E.-B.}\ \bibnamefont {Kley}}, \bibinfo {author}
  {\bibfnamefont {A.}~\bibnamefont {T\"unnermann}}, \bibinfo {author}
  {\bibfnamefont {T.}~\bibnamefont {Pertsch}}, \ and\ \bibinfo {author}
  {\bibfnamefont {F.}~\bibnamefont {Lederer}},\ }\bibfield  {title} {\enquote
  {\bibinfo {title} {Asymmetric transmission of linearly polarized light at
  optical metamaterials},}\ }\href {\doibase 10.1103/PhysRevLett.104.253902}
  {\bibfield  {journal} {\bibinfo  {journal} {Phys. Rev. Lett.}\ }\textbf
  {\bibinfo {volume} {104}},\ \bibinfo {pages} {253902} (\bibinfo {year}
  {2010})}\BibitemShut {NoStop}%
\bibitem [{\citenamefont {Faniayeu}\ \emph {et~al.}(2017)\citenamefont
  {Faniayeu}, \citenamefont {Khakhomov}, \citenamefont {Semchenko},\ and\
  \citenamefont {Mizeikis}}]{Faniayeu_APL_2017}%
  \BibitemOpen
  \bibfield  {author} {\bibinfo {author} {\bibfnamefont {I.}~\bibnamefont
  {Faniayeu}}, \bibinfo {author} {\bibfnamefont {S.}~\bibnamefont {Khakhomov}},
  \bibinfo {author} {\bibfnamefont {I.}~\bibnamefont {Semchenko}}, \ and\
  \bibinfo {author} {\bibfnamefont {V.}~\bibnamefont {Mizeikis}},\ }\bibfield
  {title} {\enquote {\bibinfo {title} {Highly transparent twist polarizer
  metasurface},}\ }\href {\doibase 10.1063/1.4994777} {\bibfield  {journal}
  {\bibinfo  {journal} {Appl. Phys. Lett.}\ }\textbf {\bibinfo {volume}
  {111}},\ \bibinfo {pages} {111108} (\bibinfo {year} {2017})}\BibitemShut
  {NoStop}%
\bibitem [{\citenamefont {Berova}, \citenamefont {Nakanishi},\ and\
  \citenamefont {Woody}(2000)}]{berova_book_2000}%
  \BibitemOpen
  \bibfield  {author} {\bibinfo {author} {\bibfnamefont {N.}~\bibnamefont
  {Berova}}, \bibinfo {author} {\bibfnamefont {K.}~\bibnamefont {Nakanishi}}, \
  and\ \bibinfo {author} {\bibfnamefont {R.~W.}\ \bibnamefont {Woody}},\
  }\href@noop {} {\emph {\bibinfo {title} {Circular Dichroism. Principles and
  Applications}}}\ (\bibinfo  {publisher} {Wiley, New York},\ \bibinfo {year}
  {2000})\BibitemShut {NoStop}%
\bibitem [{\citenamefont {Collett}(1993)}]{Collett_1993}%
  \BibitemOpen
  \bibfield  {author} {\bibinfo {author} {\bibfnamefont {E.}~\bibnamefont
  {Collett}},\ }\href@noop {} {\emph {\bibinfo {title} {Polarized Light:
  Fundamentals and Applications}}}\ (\bibinfo  {publisher} {Dekker, New York},\
  \bibinfo {year} {1993})\BibitemShut {NoStop}%
\end{thebibliography}%

\end{document}